\documentclass[aps,prd,preprint,groupedaddress,showpacs]{revtex4}

\usepackage{epsfig}
\usepackage{graphics}
\usepackage{slashed}
\usepackage{color}
\usepackage[citecolor=true]{hyperref}
\usepackage{multirow}
\usepackage{amsmath}
\begin{document}

\leftmargin -2cm
\def\choosen{\atopwithdelims..}

\boldmath
\title{Associated production of heavy quarkonium and $D-$meson in the improved color evaporation model
with KaTie} \unboldmath

\author{\firstname{A.A.}\surname{Chernyshev}} \email{aachernyshoff@gmail.com}

\affiliation{Samara National Research University, Moskovskoe Shosse,
34, 443086, Samara, Russia}

\author{\firstname{V.A.}\surname{Saleev}} \email{saleev.vladimir@gmail.com}

\affiliation{Samara National Research University, Moskovskoe Shosse,
34, 443086, Samara, Russia}

\affiliation{Joint Institute for Nuclear Research, Dubna, 141980
Russia}

\begin{abstract}
In the article, we study associated production of prompt
$J/\psi(\Upsilon)$  and $D-$mesons in the improved color evaporation
model using the high-energy factorization approach as it is realized
in the Monte-Carlo event generator KaTie. The modified
Kimber-Martin-Ryskin-Watt model for unintegrated parton distribution
functions is used. We predict cross sections for associated
$J/\psi(\Upsilon)$ and $D-$meson hadroproduction via the single and
double parton scattering mechanisms using the set of model
parameters which has been fixed early for description of prompt
single and pair heavy quarkonium production at the LHC energies. We
found the results of calculations agree with the LHCb Collaboration
data at the energies $\sqrt{s} = 7,8$ TeV and we present theoretical
predictions for the energy $\sqrt{s} = 13$ TeV . \vspace{0.2cm}
\end{abstract}


\maketitle

\section{Introduction}
\label{sec:introduction}

Measurements of the $J/\psi+D$ and $\Upsilon+D$ (here and below
$\Upsilon=\Upsilon(1S)$) associated production by the LHCb
Collaboration \cite{LHCb:2012aiv, LHCb:2015wvu} at the energies
$\sqrt{s}=7,8$ TeV clear demonstrate a dominant role of the double
parton scattering (DPS) mechanism of the parton model compared with
the conventional single parton scattering (SPS) scenario. The
average value of the DPS parameter $\sigma_{eff}$
 extracted in the $\Upsilon+D$ pair production
is about $\sigma_{eff}=18.0 \pm 1.8$ mb and in the $J/\psi+D$ it is
about $\sigma_{eff}=15.8 \pm 2.4$ mb~ \cite{LHCb:2012aiv,
LHCb:2015wvu}. For today, the extractions of parameter $\sigma_{\rm
eff}$, based on the DPS pocket formula \cite{Calucci:1999yz}, have
been obtained in different experiments. Values of $\sigma_{\rm
eff}=2-25$ mb have been derived, though with large errors, with a
simple average giving $\sigma_{\rm eff}=15$ mb
~\cite{Chapon:2020heu}.

 Theoretical calculations performed in the leading order (LO) in $\alpha_S$ of the
collinear parton model within the Color Singlet Model
\cite{CSM1,CSM2} and within the approach of the Nonrelativistic
Quantum Chromodynamics (NRQCD)~\cite{NRQCD}  predict very small
values of the SPS cross sections for $J/\psi+D$ \cite{Shao:2020kgj}
and $\Upsilon+D$ pair production
\cite{Berezhnoy:2015jga,Likhoded:2015fdr}. In the
Ref.~\cite{Karpishkov:2019vyt}, the associated $\Upsilon+D$ pair
production was studied in the $k_T-$factorization
\cite{KT1,KT2,Gribov} and  NRQCD and it was demonstrated that SPS
contribution to the cross section is also sufficiently smaller the
experimental data from LHCb collaboration \cite{LHCb:2015wvu}. The
new LHCb measurements of $J/\psi +J/\psi$ and $J/\psi +\Upsilon$
pair production cross sections~\cite{LHCb:2023qgu,LHCb:2023ybt}
motivate to make predictions for $J/\psi+D$ and $\Upsilon+D$ pair
production at the $\sqrt{s}=7$ and $13$ TeV. Instead of previous
theoretical studies for such processes, we use Improved Color
Evaporation Model (ICEM) \cite{ICEM_Ma_Vogt} to describe
hadronization of heavy quark and antiquark pair into a final
quarkonium. The ICEM was successfully used recently to describe
single $J/\psi (\Upsilon)$ production both in the collinear parton
model \cite{ICEM2017,ICEM2021} and in the $k_T-$factorization
\cite{ICEM_KT1,ICEM_KT2}. The pair quarkonium production in the ICEM
using the $k_T-$factorization was studied in Refs.
\cite{ChernyshevSaleev2Psi2022,ChernyshevSaleev2Ups2022}. The
D-meson production at the LHC energies was described in the
$k_T-$factorization and the fragmentation approach using $c\to D$
nonperturbative fragmentation function $D_{c\to D}(z)$ in
Refs.\cite{Maciula:2016wci,vanHameren:2015wva}.

In the study, we calculate cross section for associated $J/\psi+D$
and $\Upsilon+D$ production in the proton-proton collisions in the
$k_T$-factorization~\cite{KT1,KT2,Gribov} using the Monte-Carlo (MC)
event generator KaTie \cite{katie}. Following by the parton
Reggeization approach
(PRA)~\cite{Karpishkov:2017kph,NefedovSaleev2020}, which is a
gauge-invariant version of the $k_T-$factorization, the modified
Kimber-Martin-Ryskin-Watt model~ \cite{KMR,WMR} for unintegrated
parton distribution functions (uPDFs) is
used~\cite{NefedovSaleev2020}.

\section{  Event generator KaTie and unintegrated parton distribution functions }
\label{sec:KaTie}

We apply fully numerical method of the calculation using the parton
level event generator KaTie~\cite{katie}. The approach to obtaining
gauge invariant amplitudes with off-shell initial state partons in
scattering at high-energy multi-Regge kinematics was proposed in the
Ref.~\cite{hameren1,hameren2}. The method is based on the use of
spinor amplitudes formalism and recurrence relations of the
Britto-Cachazo-Feng-Witten (BCFW) type. This
formalism~\cite{katie,hameren1,hameren2} for numerical amplitude
generation is equivalent to amplitudes built according to Feynman
rules of the Lipatov Effective Field Theory at the level of tree
diagrams~\cite{NSS2013,kutak}. The accuracy of numerical
calculations using KaTie for total proton-proton cross sections is
taking as 0.1 \%.

 In the high-energy factorization or
$k_T-$factorization, the cross section for the hard process $p + p
\to {\cal Q} + X$ in the multi-Regge kinematics is calculated as
integral convolution of the parton cross section $d\hat\sigma(i + j
\to {\cal Q}+X)$ and the unintegrated parton distribution functions
(uPDFs) by the factorization formula
\begin{eqnarray}
  d\sigma & = & \sum_{i, \bar{j}}
    \int\limits_{0}^{1} \frac{dx_1}{x_1} \int \frac{d^2{\bf q}_{T1}}{\pi}
{\Phi}_i(x_1,t_1,\mu^2)
    \int\limits_{0}^{1} \frac{dx_2}{x_2} \int \frac{d^2{\bf q}_{T2}}{\pi}
{\Phi}_{j}(x_2,t_2,\mu^2)\cdot d\hat{\sigma}, \label{eqI:kT_fact}
\end{eqnarray}
where $t_{1,2} = - {\bf{q}}_{1,2T}^2$,
$q_{1,2}^\mu=x_{1,2}P_{1,2}^\mu+ q_{1,2T}^\mu$,
$q_{1,2T}=(0,{\bf{q}}_{1,2T},0)$, the cross section of the
subprocess with off-shell initial partons, which are treated as
Reggeized partons~\cite{Lipatov95,LipatovVyazovsky}, $d\hat{\sigma}$
is expressed in terms of squared Reggeized amplitudes
$\overline{|{\mathcal{A}}_{\mathrm{PRA}}|^2}$
 in a standard way~\cite{NSS2013}.

The unPDFs can be written as follows from the KMRW model
\cite{KMR,WMR}:
\begin{equation}
\Phi_i(x,t,\mu^2)= \frac{\alpha_s(\mu)}{2\pi}
\frac{T_i(t,\mu^2,x)}{t} \sum\limits_{j=q,\bar{q},g}\int\limits_x^1
dz\ P_{ij}(z) {F}_j\left( \frac{x}{z}, t \right) \theta\left(
\Delta(t,\mu^2)-z \right),\label{uPDF}
\end{equation}
where $F_i(x,\mu^2)=x f_j(x,\mu^2)$. To resolve infra-red
divergence, the following cutoff on $z_{1,2}$ can be derived:
$z_{1,2}< 1-\Delta_{KMR}(t_{1,2},\mu^2),$ where
$\Delta_{KMR}(t,\mu^2)=\sqrt{t}/(\sqrt{\mu^2}+\sqrt{t})$ is the
KMR-cutoff function~\cite{KMR}. To resolve collinear divergence
problem, we require that modified uPDF ${\Phi}_i(x,t,\mu^2)$ should
be satisfied exact normalization condition:
\begin{equation}
\int\limits_0^{\mu} dt \Phi_i(x,t,\mu^2) = {F}_i(x,\mu^2),
\end{equation}
which is equivalent to:
\begin{equation}
\Phi_i(x,t,\mu^2)=\frac{d}{dt}\left[ T_i(t,\mu^2,x){F}_i(x,t)
\right],\label{eq:sudakov}
\end{equation}
where $T_i(t,\mu^2,x)$  is referred to as Sudakov form-factor,
satisfying the boundary conditions $T_i(t=0,\mu^2,x)=0$ and
$T_i(t=\mu^2,\mu^2,x)=1$.

The solution for Sudakov form-factor in Eq. ({\ref{eq:sudakov}) has
been obtained in Ref.~\cite{NefedovSaleev2020}:
\begin{equation}
T_i(t,\mu^2,x)=\exp\left[ -\int\limits_t^{\mu^2} \frac{dt'}{t'}
\frac{\alpha_s(t')}{2\pi} \left( \tau_i(t',\mu^2) + \Delta\tau_i
(t',\mu^2,x) \right) \right]\label{eq:sud}
\end{equation}
with
\begin{eqnarray*}
\tau_i(t,\mu^2)&=&\sum\limits_j \int\limits_0^1 dz\ zP_{ji}(z)\theta (\Delta(t,\mu^2)-z), \label{eq:tau} \\
\Delta\tau_i(t,\mu^2,x)&=& \sum\limits_j \int\limits_0^1 dz\
\theta(z-\Delta(t,\mu^2)) \left[ zP_{ji}(z) -
\frac{{F}_j\left(\frac{x}{z},t \right)}{{F}_i(x,t)} P_{ij}(z)
\theta(z-x) \right].
\end{eqnarray*}
In our modified KMRW model, the Sudakov form-factor (\ref{eq:sud})
contains the $x-$depended $\Delta \tau_i$-term in the exponent which
is needed to preserve exact normalization condition for arbitrary
$x$ and $\mu$. There is a numerically-important difference that in
our uPDFs the rapidity-ordering condition is imposed both on quarks
and gluons, while in KMRW approach it is imposed only on gluons.

\section{ICEM }
\label{sec:ICEM}

In the ICEM, the cross section for the production of  heavy
quarkonium ${\cal Q} = J/\psi, \Upsilon$ is related to the cross
section for the production of $q \bar{q}-$pair as follows $(q=c,b)$:
\begin{equation}
\sigma(p + p \to \mathcal{Q} + X) = \mathcal{F}^{\mathcal{Q}} \times
\int\limits_{m_{\mathcal{Q}}}^{2m_{D,B}} \frac{d\sigma(p + p \to q +
\bar{q} + X)}{dM} dM,
\end{equation}
where $M$ is the invariant mass of the $q \bar{q}$-pair with
$4$-momentum $p^{\mu}_{q \bar{q}} = p^{\mu}_{q} +
p^{\mu}_{\bar{q}}$, $m_{\mathcal{Q}}$ is the mass of the quarkonium,
and $m_{D,B}$ are the masses of the lightest  $D$ and $B$ mesons.
Parameter $\mathcal{F}^{\mathcal{Q}}$ is considered as a probability
of transformation of the $q \bar{q}$-pair with invariant mass
$m_{\mathcal{Q}} < M < 2 m_{D,B}$ into the quarkonium $\mathcal{Q}$.

The cross section for the associated production of quarkonium
$\mathcal{Q}$ and $D-$meson in the ICEM is related to the cross
section for the associated production of $q \bar{q}$-pair and
$D-$meson in the SPS as follows:
\begin{equation}
\sigma^{\rm SPS}(p + p \to \mathcal{Q} + D + X) =
\mathcal{F}^{\mathcal{Q}} \times
\int\limits_{m_{\mathcal{Q}}}^{2m_{D,B}} \frac{d\sigma(p + p \to q +
\bar{q} + D + X)}{dM} dM.
\end{equation}

\begin{figure}[h]
\begin{center}
\includegraphics[width=0.8\textwidth,angle=0]{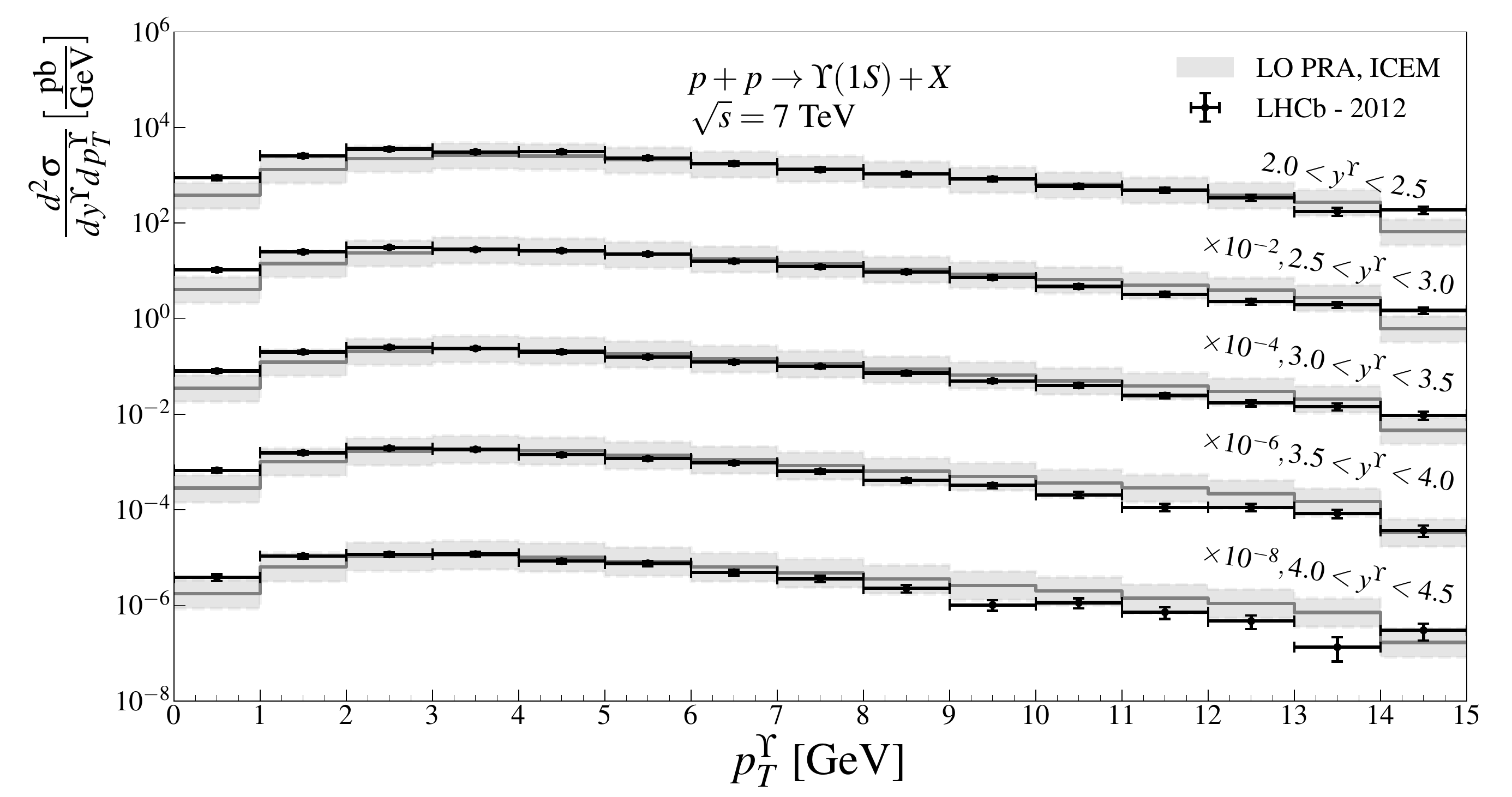}
\vspace{-3mm} \caption{ Prompt $\Upsilon(1S)$ production cross
section as a function of transverse momenta at the $\sqrt{s}=7$ TeV
at the different ranges of rapidity. The data are from LHCb
Collaboration \cite{LHCb:2015log}. \label{fig_1} }
\end{center}
\end{figure}

\begin{figure}[h]
\begin{center}
\includegraphics[width=0.8\textwidth,angle=0]{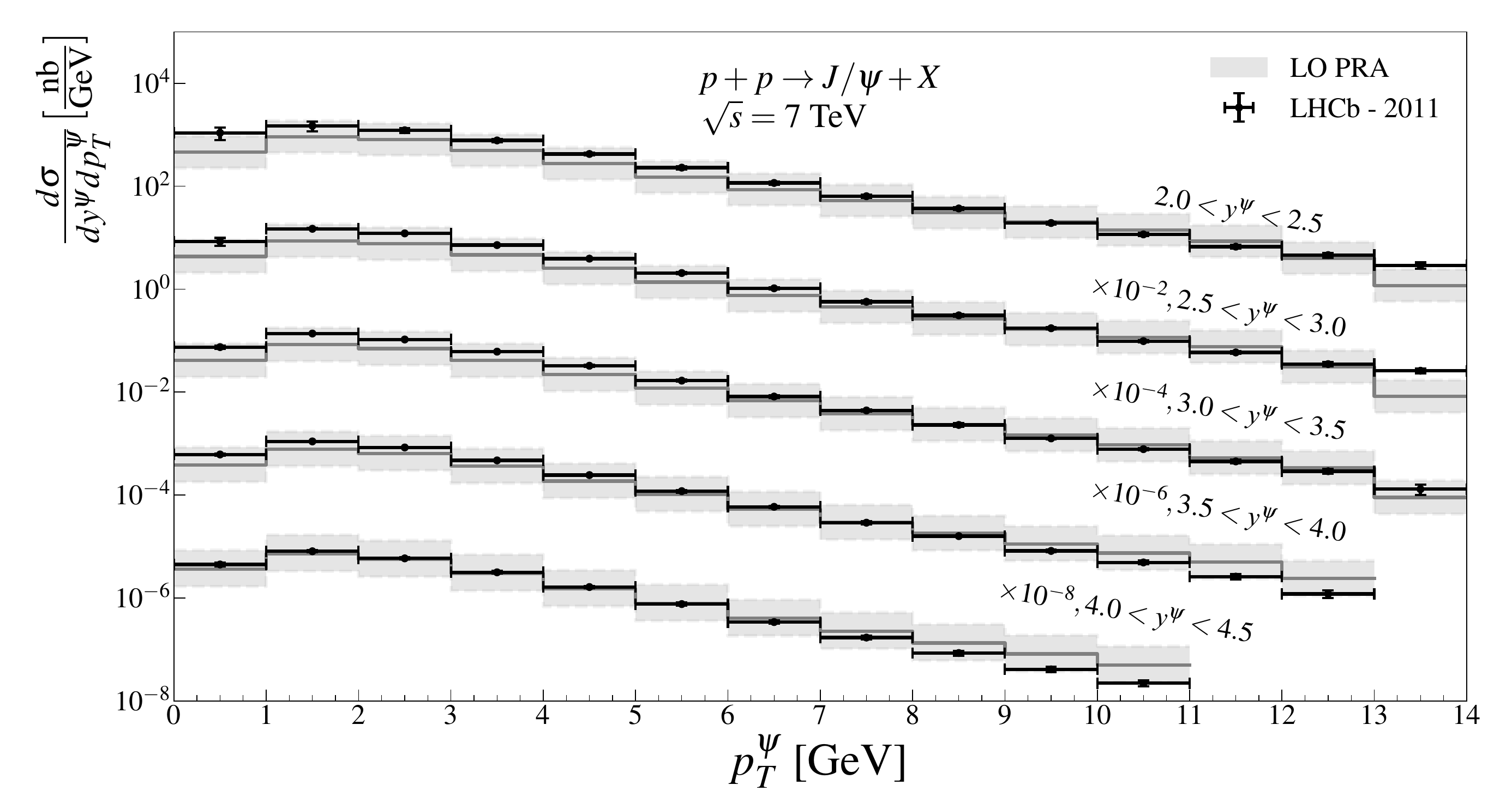}
\vspace{-3mm} \caption{ Prompt $J/\psi$ production cross section as
a function of transverse momenta at the $\sqrt{s}=7$ TeV at the
different ranges of rapidity. The data are from LHCb Collaboration
\cite{LHCb:2011zfl}. \label{fig_2} }
\end{center}
\end{figure}

In the DPS scenario, the cross section for the associated production
of a quarkonium $\mathcal{Q}$ and $D-$meson is expressed in terms of
the cross sections of two independent subprocesses
\begin{equation}
\sigma^{\rm DPS}(p + p \to \mathcal{Q} + D + X) = \frac{ \sigma^{\rm
SPS}(p + p \to \mathcal{Q} + X_1) \times \sigma^{\rm SPS}(p + p \to
D + X_2) } {\sigma_{\rm eff}},
\end{equation}
where parameter $\sigma_{\rm eff}$  controls the contribution of the
DPS mechanism. To calculate $J/\psi + D$ and $\Upsilon + D$
associated productions we take here  parameters ${\cal
F}^{\psi}\simeq {\cal F}^{\Upsilon} = 0.02$ as it was early obtained
by the fit of the LHCb data for the single prompt $J/\psi$
production cross section~ \cite{ChernyshevSaleev2Psi2022} and single
prompt $\Upsilon$ production~\cite{ChernyshevSaleev2Ups2022} using
the ICEM and event generator KaTie. The DPS parameter is taken equal
$\sigma_{\rm eff} = 11$ mb as it follows from the fit of $J/\psi$
and $\Upsilon$ pair production cross sections and spectra using the
same
approaches~\cite{ChernyshevSaleev2Psi2022,ChernyshevSaleev2Ups2022}.

In the Figs.~\ref{fig_1} and \ref{fig_2}, we plot theoretical
predictions for transverse momentum spectra $(p_T)$ of prompt
$\Upsilon$ and $J/\psi$ obtained within the ICEM and
$k_T-$factorization using generator KaTie~\cite{katie} and modified
uPDFs~\cite{NefedovSaleev2020}. The good agreement with the LHCb
data \cite{LHCb:2015log,LHCb:2011zfl} at the $\sqrt{s}=7$  is
founded. The shadow bounds demonstrate theoretical uncertainty
following from the choice of the hard scale $\mu$, which is taken as
$\mu=\xi \sqrt{M_{Q}^2+p_T^2}$ with $\xi=\frac{1}{2},~1,~2$.

\section{Fragmentation approach}
\label{sec:Fragmentation}
 For description of the inclusive
production of an open charm meson it is often used the fragmentation
approach in which the cross section for the production of $D-$meson
is related to the $c \bar c$ pair production by the following way:
\begin{equation}
 \frac{d \sigma^{\rm SPS}(p + p \to D + X)}{d^2 p_{TD}dy_D} = \int
\limits_{z_{\rm cut}}^{1} dz \ {\cal D}_{c \to D}(z) \ \frac{d
\sigma^{\rm SPS}(p + p \to c + \bar c + X)}{d^2 p_{Tc}dy_c},
\end{equation}
where ${\cal D}_{c \to D}(z)$ is a fragmentation function (FF) of
the $c-$quark into $D-$meson, $p_c$ is $c-$quark 4-momentum
expressed in terms of $D-$meson 4-momentum $p_D$ through parameter
$z$, which is defined as
\begin{equation*}
z = \frac{E_D + |{\bf p}_D|}{E_c + |{\bf p}_c|}.
\end{equation*}
The minimal value of parameter   $z_{\rm cut} = m_D / (E_c + |{\bf
p}_c|)$  cuts out the non-physical region, where $E_c < m_D$, and we
apply collinear fragmentation approximation,
 ${{\bf p}_D}/{|{\bf
p}_D|}={{\bf p}_c}/{|{\bf p}_c|}$. In our calculations, we use
so-called Peterson FF
\begin{equation}
{\cal D}_{c \to D}(z) = {\cal N} \ \frac{z \ (1 - z)^2}{[(1 - z)^2 +
\epsilon \, z]^2}
\end{equation}
with parameter $\epsilon = 0.06$. FF is normalized such that
\begin{equation}
\int_0^1 dz \ {\cal D}_{c \to D}(z) = P_{c\to D},
\end{equation}
where $P_{c\to D^+}=0.225$ and $P_{c\to D^0}=0.542$.
\cite{Gladilin:1999nf}.

To test fragmentation approach, we performed calculations for
$D^{+,0}-$meson transverse momentum spectra and compare obtained
results with the relevant LHCb data \cite{LHCb:2013xam} at the
energy $\sqrt{s}=7$ TeV, as it is shown in the Figs.~\ref{fig_3} and
\ref{fig_4}.

Such a way, we demonstrate applicability of our approach based on
the $k_T-$factorization with the modified KMRW unPDFs, the ICEM and
the fragmentation model, for description of single heavy quarkonium
and single D-meson production. Now, we are in position to study
associated $J/\psi(\Upsilon)+D$ production.

\begin{figure}[h]
\begin{center}
\includegraphics[width=0.8\textwidth,angle=0]{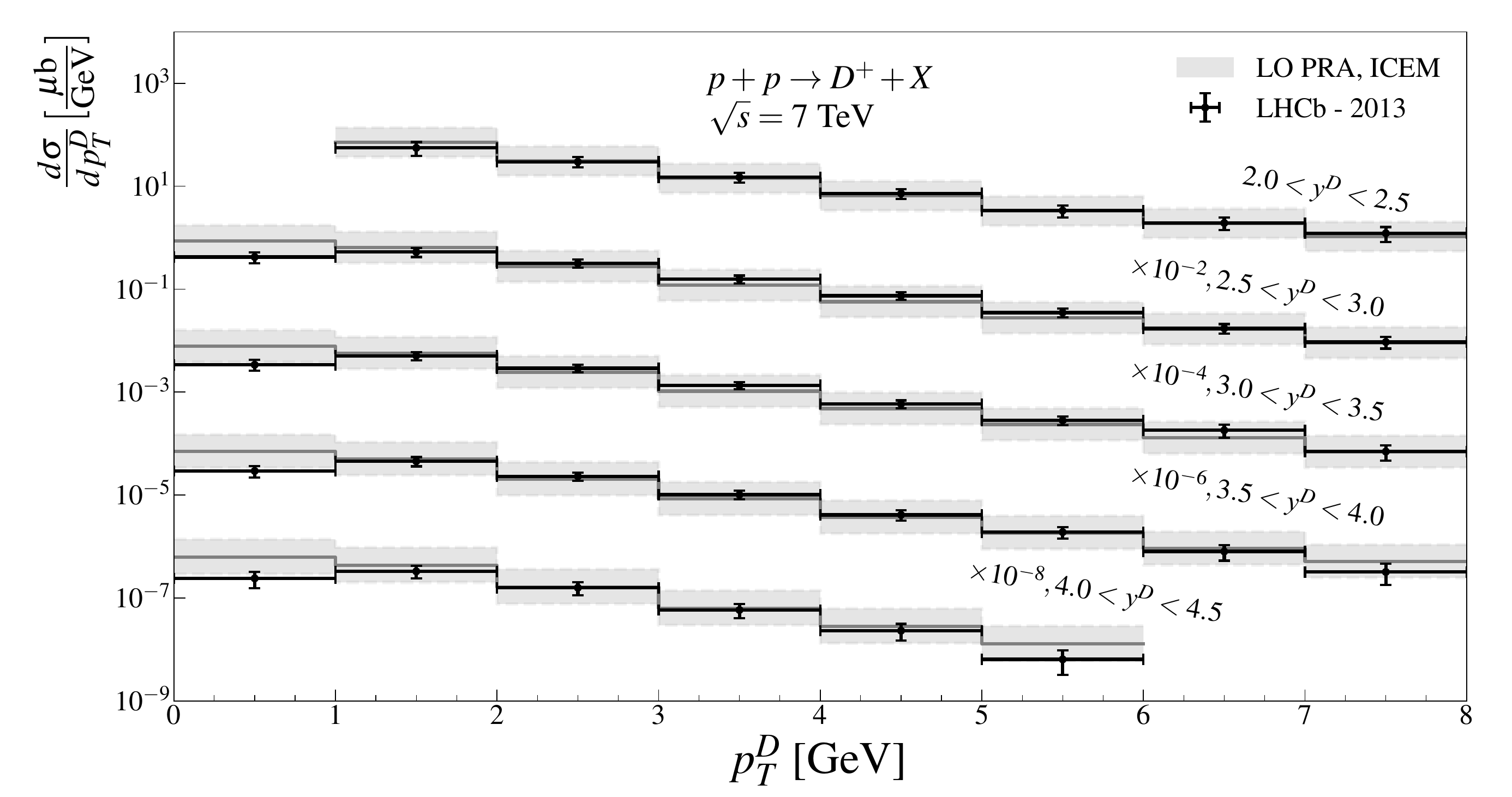}
\vspace{-3mm} \caption{ $D^+$-meson production cross sections as
functions of transverse momenta at the $\sqrt{s}=7$ TeV. The data
are from the LHCb Collaboration \cite{LHCb:2013xam}. \label{fig_3} }
\end{center}
\end{figure}

\begin{figure}[h]
\begin{center}
\includegraphics[width=0.8\textwidth,angle=0]{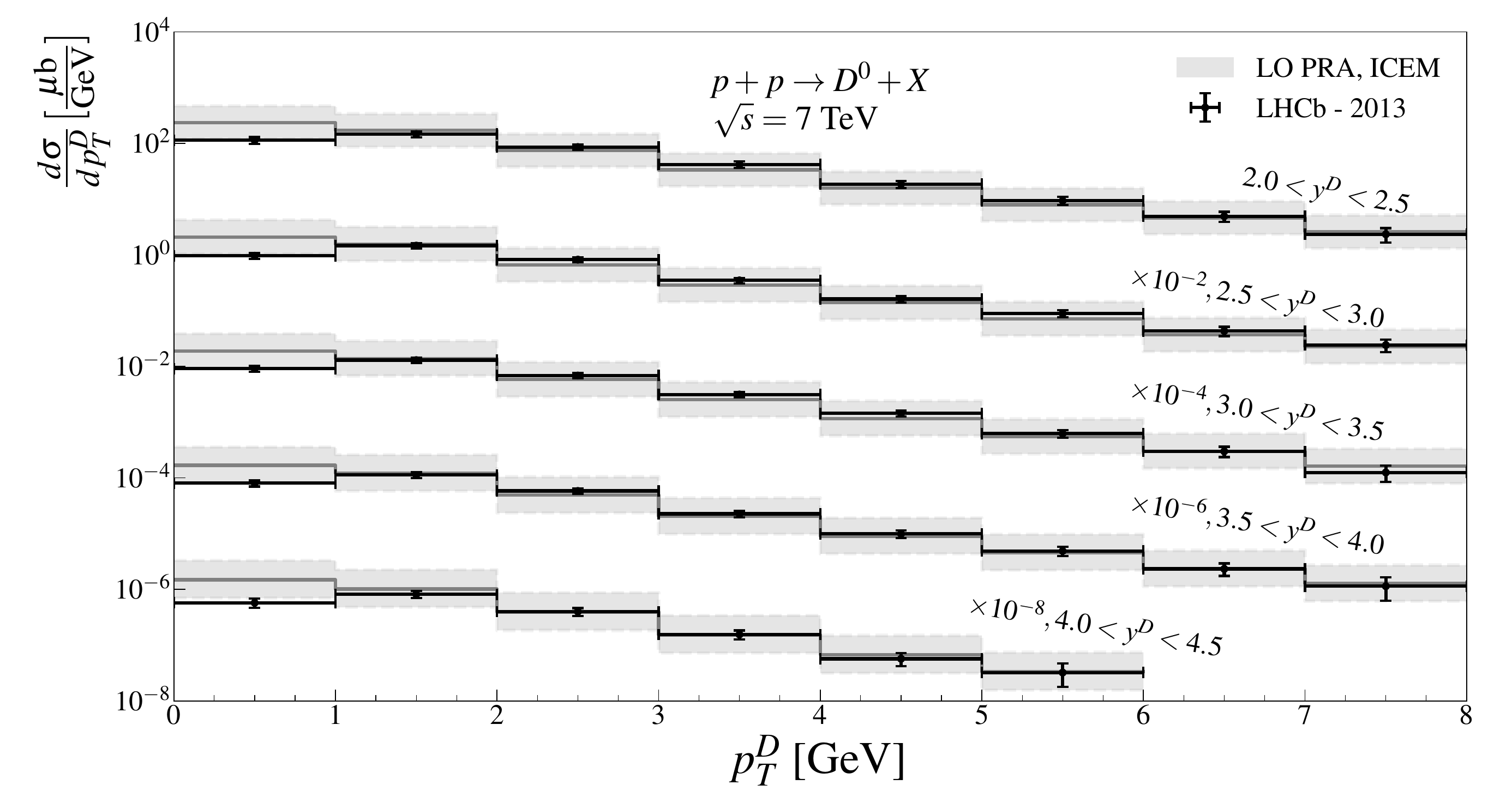}
\vspace{-3mm} \caption{ $D^0$-meson production cross sections as
functions of transverse momenta at the $\sqrt{s}=7$ TeV. The data
are from the LHCb Collaboration \cite{LHCb:2013xam}. \label{fig_4} }
\end{center}
\end{figure}

\section{Associated $J/\psi(\Upsilon)+D$ production}

\label{sec:results} In case of $J/\psi + D$ associated production
via the SPS, we take into account contributions of the following
parton subprocesses:
\begin{align}
R + R & \to c + \bar c + c + \bar c, \\
Q_q + \bar{Q}_q & \to c + \bar c + c + \bar c,
\end{align}
where $R$ is a Reggeized gluon, $Q_q (\bar{Q}_q)$ as a Reggeized
quark (antiquark), and $q = u, d, s, c$. In the DPS approach, the
processes of the $J/\psi + D$ associated production are  following:
\begin{align}
R + R & \to c + \bar c, \\
Q_q + \bar{Q}_q & \to c + \bar c
\end{align}
in both groups.

In case of $\Upsilon + D$ associated production via the SPS, we take
into account contributions of the following parton subprocesses:
\begin{align}
R + R & \to b + \bar b + c + \bar c, \\
Q_q + \bar{Q}_q & \to b + \bar b + c + \bar c.
\end{align}
In the DPS approach, the processes of $\Upsilon + D$ associated
production are following:
\begin{align}
R + R & \to b + \bar b, \\
Q_q + \bar{Q}_q & \to b + \bar b \\
R + R & \to c + \bar c, \\
Q_q + \bar{Q}_q & \to c + \bar c.
\end{align}
Using the  KaTie, we can do calculations up to four particles in a
final state that it is enough for our purposes. We put masses of
quarks, heavy mesons and heavy quarkonia during the calculations
equal $m_c=1.3$ GeV, $m_b=4.5$ GeV, $m_{D^+}=1.87$ GeV,
$m_{D^0}=1.86$ GeV, $m_{B^+}=5.28$ GeV, $m_{J/\psi}=3.097$ GeV,
$m_{\Upsilon}=9.46$ GeV. As it was already mentioned in the Section
\ref{sec:KaTie}, we use modified KMRW uPDFs
\cite{NefedovSaleev2020}, which were used previously in our
calculations for $J/\psi(\Upsilon)$ pair production  using event
generator KaTie
\cite{ChernyshevSaleev2Psi2022,ChernyshevSaleev2Ups2022}.

The results of our calculations for associated $J/\psi+D^{+,0}$
production are presented in the Figs. \ref{fig_5} and \ref{fig_6}
and for associated $\Upsilon+D^{+,0}$ production in the Figs.
\ref{fig_7} and \ref{fig_8}. We have obtained quite satisfactory
agreement between data and our calculations for all spectra in
$J/\psi+D$ and $\Upsilon +D$ associated production. There are
disagreements only in the rapidity difference spectra, especially in
$J/\psi+D$ production, at the large values of $|\Delta y^{\psi D}|$.
We find theoretical calculations overestimate LHCb data at the
$|\Delta y^{\psi D}|\simeq 1.75-2.0$ about one order of magnitude.

In the Table \ref{Table:1} we collect our theoretical predictions
for associated  ${J/\psi(\Upsilon)+D}$ production cross sections at
the $\sqrt{s}=7,8$ TeV and $\sqrt{s}=13$ TeV. The predictions at the
$\sqrt{s}=7,8$ TeV are compared with the experimental data. We find
a quite good agreement between data and theoretical calculations,
which correspond the default choice of the hard scale $\mu =
\frac{1}{2}\left( m^{\cal Q}_T + m^{D}_T \right)$, where $m^{\cal
Q}_T=\sqrt{m_{\cal Q}^2+p_{T,{\cal Q}}^2}$ and
$m^{D}_T=\sqrt{m_{D}^2+p_{T,{D}}^2}$. The variation of hard scale by
factor 2 around the default value gives an estimation of uncertainty
of $k_T-$factorization calculation. In case of the DPS production,
such uncertainty may be very large, about 100 \% at the up limit,
instead off the SPS production mechanism. It is due to we have the
product of four uPDFs in the DPS calculation, each of them
sufficiently  depend  on the choice of hard scale $\mu$.

\begin{figure}[h]
\begin{center}
\includegraphics[width=0.4\textwidth,angle=0]{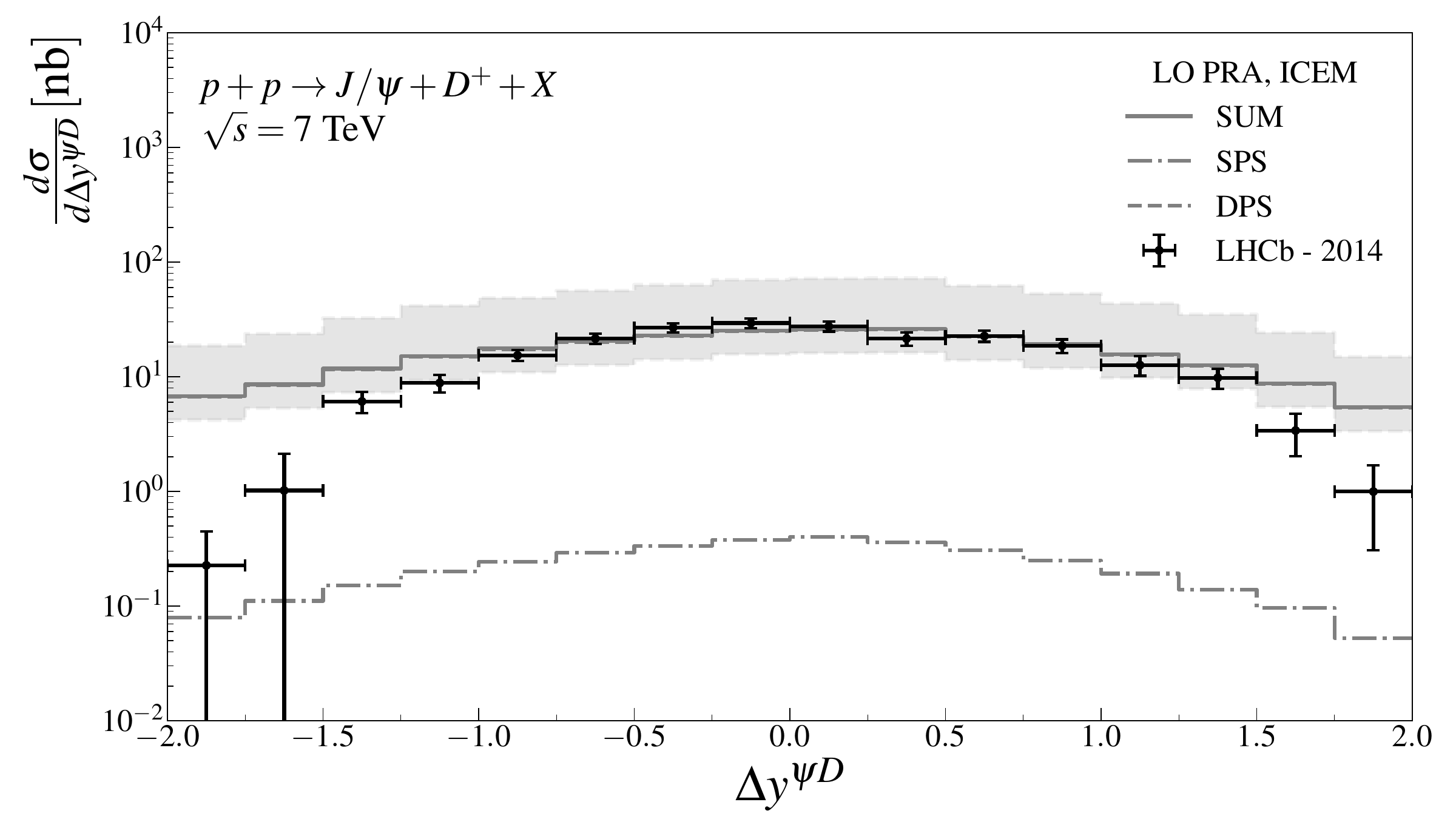}\includegraphics[width=0.4\textwidth,angle=0]{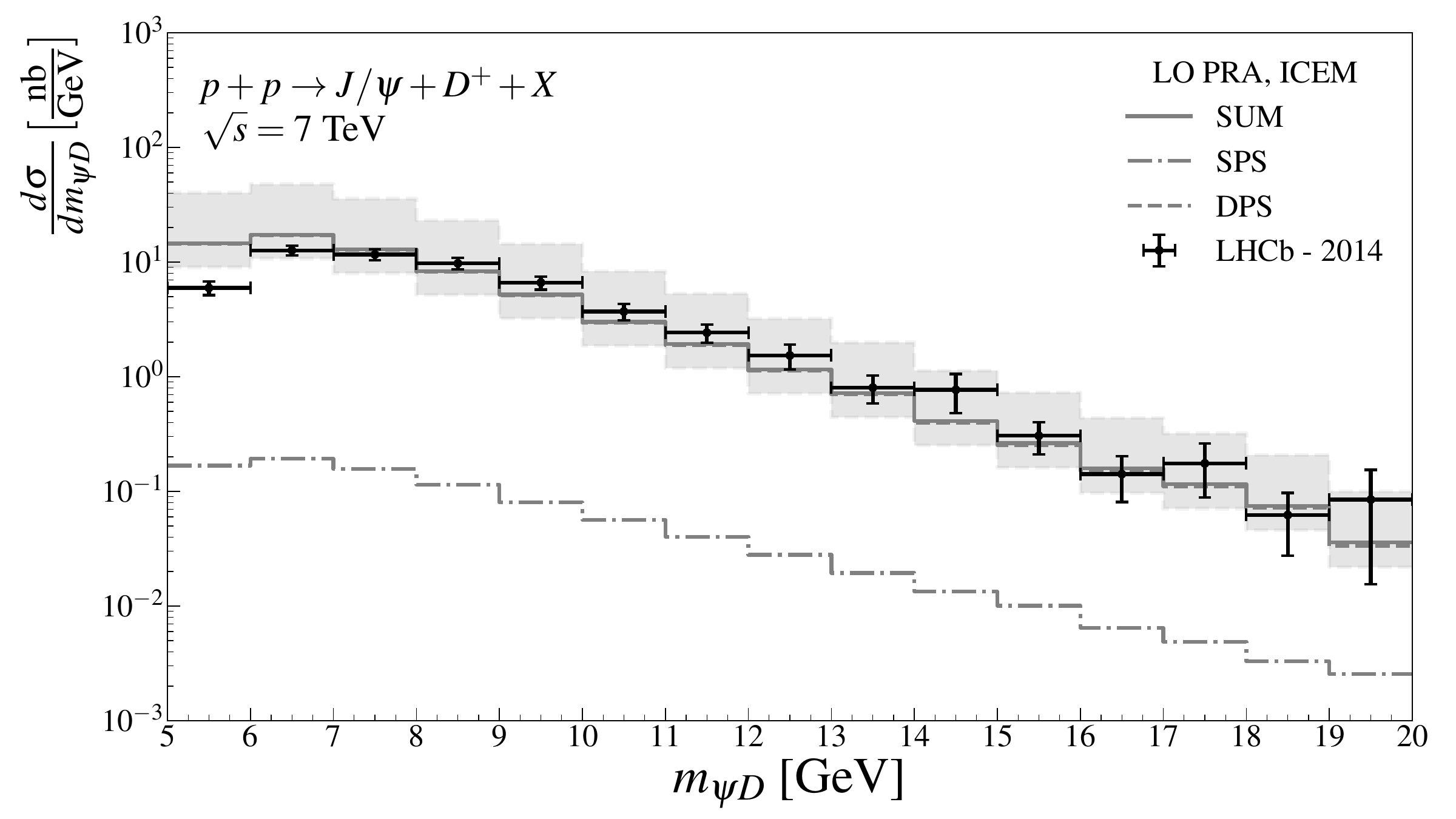}
\includegraphics[width=0.4\textwidth,angle=0]{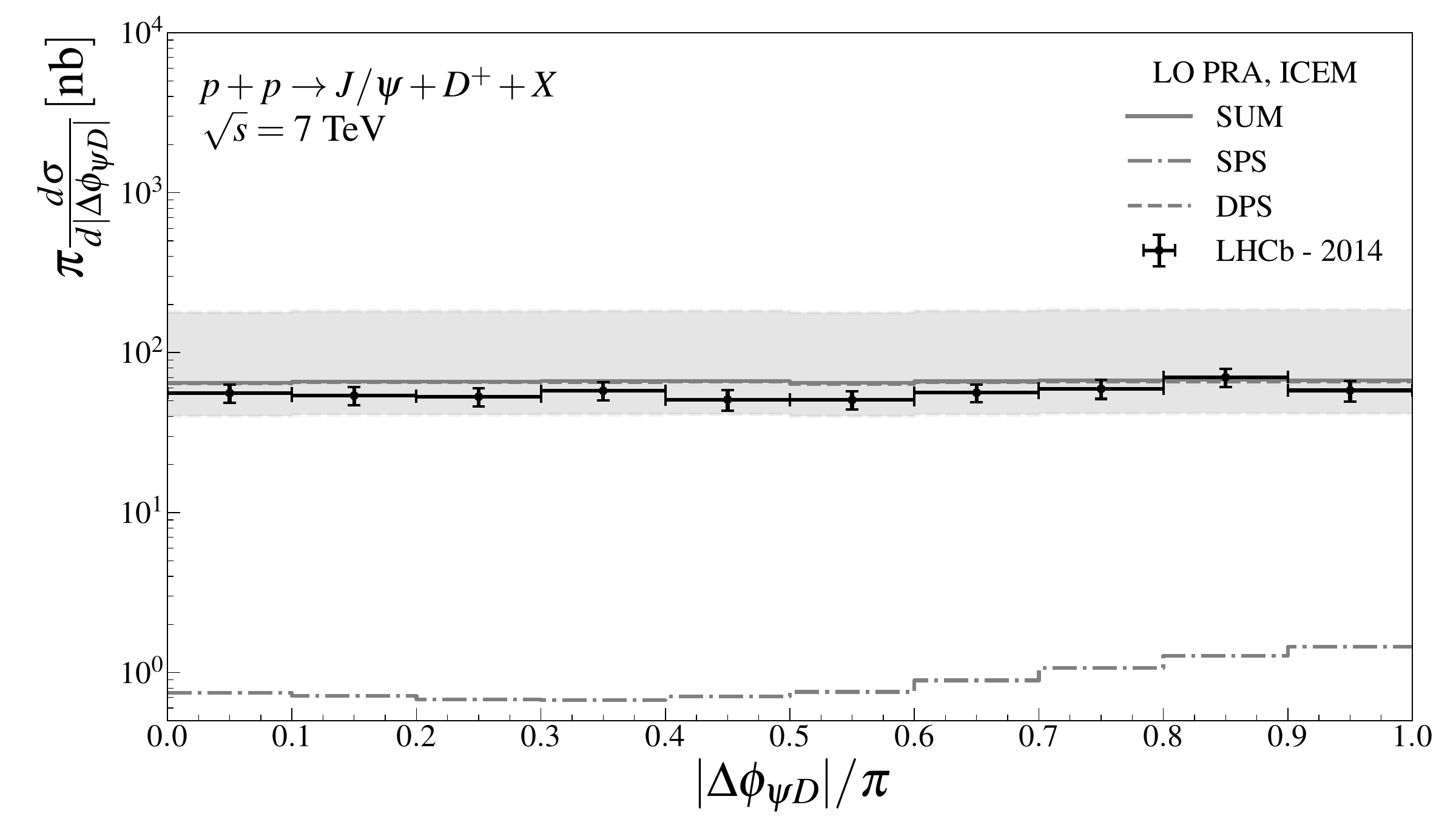}\includegraphics[width=0.4\textwidth,angle=0]{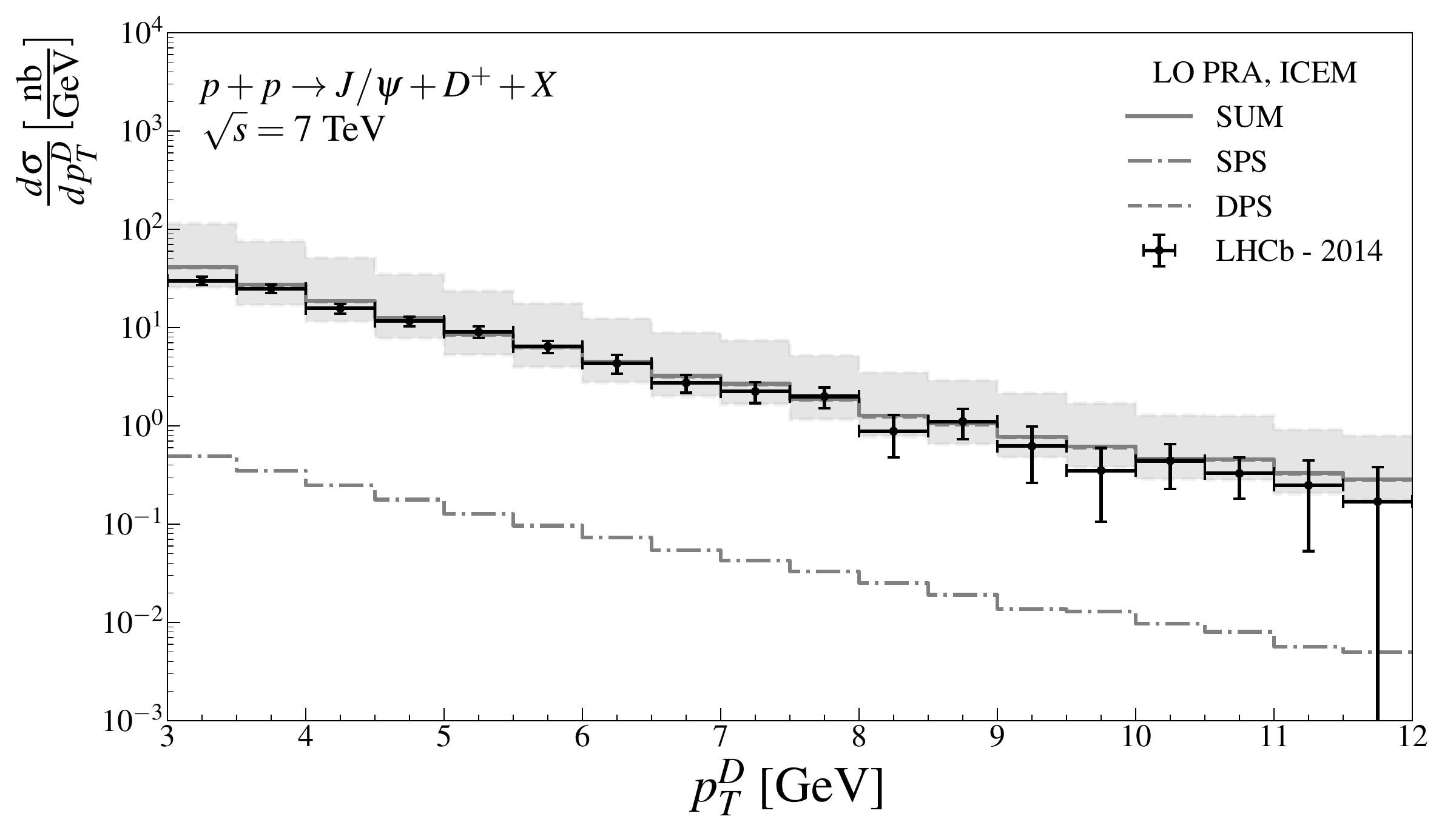}
\includegraphics[width=0.4\textwidth,angle=0]{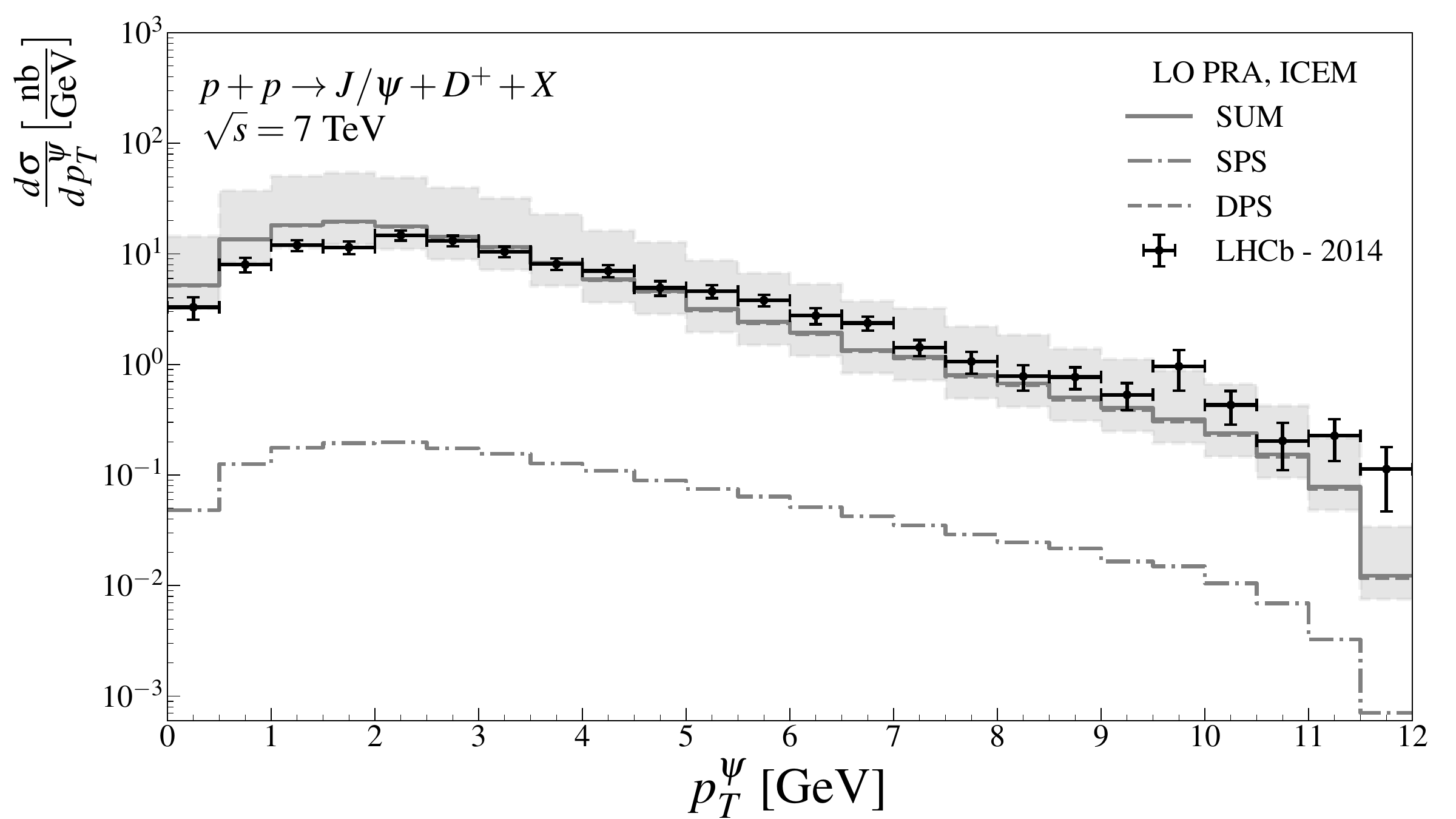}
\vspace{-3mm} \caption{ Various spectra for associated $J/\psi+D^+$
production at the $\sqrt{s}=7$ TeV: $\Delta y^{\psi D}$ - rapidity
difference, $m^{\psi D}$ - invariant mass, $|\Delta \phi^{\psi D}|$
- azimuthal angle difference, $p_T^D$ - $D^+$ transverse momentum
and $p_T^\psi$ - $J/\psi$ transverse momentum. The dashed line is
the DPS contribution, the dashed-dotted line is the SPS
contribution, solid line is they sum. Grey bounds around the solid
line are scale uncertainties of calculations. The data are from LHCb
collaboration \cite{LHCb:2012aiv}. \label{fig_5} }
\end{center}
\end{figure}

\begin{figure}[h]
\begin{center}
\includegraphics[width=0.4\textwidth,angle=0]{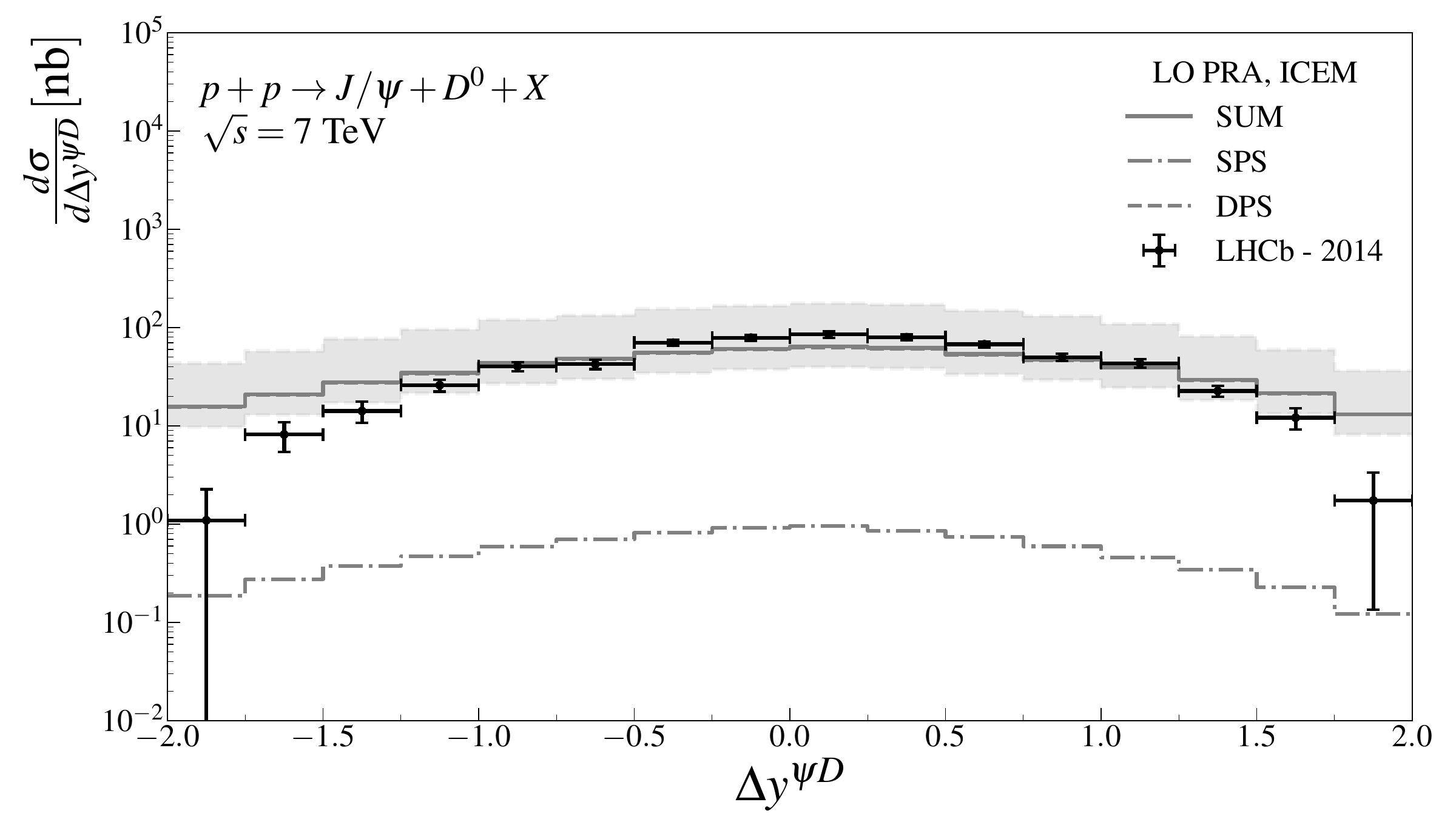}\includegraphics[width=0.4\textwidth,angle=0]{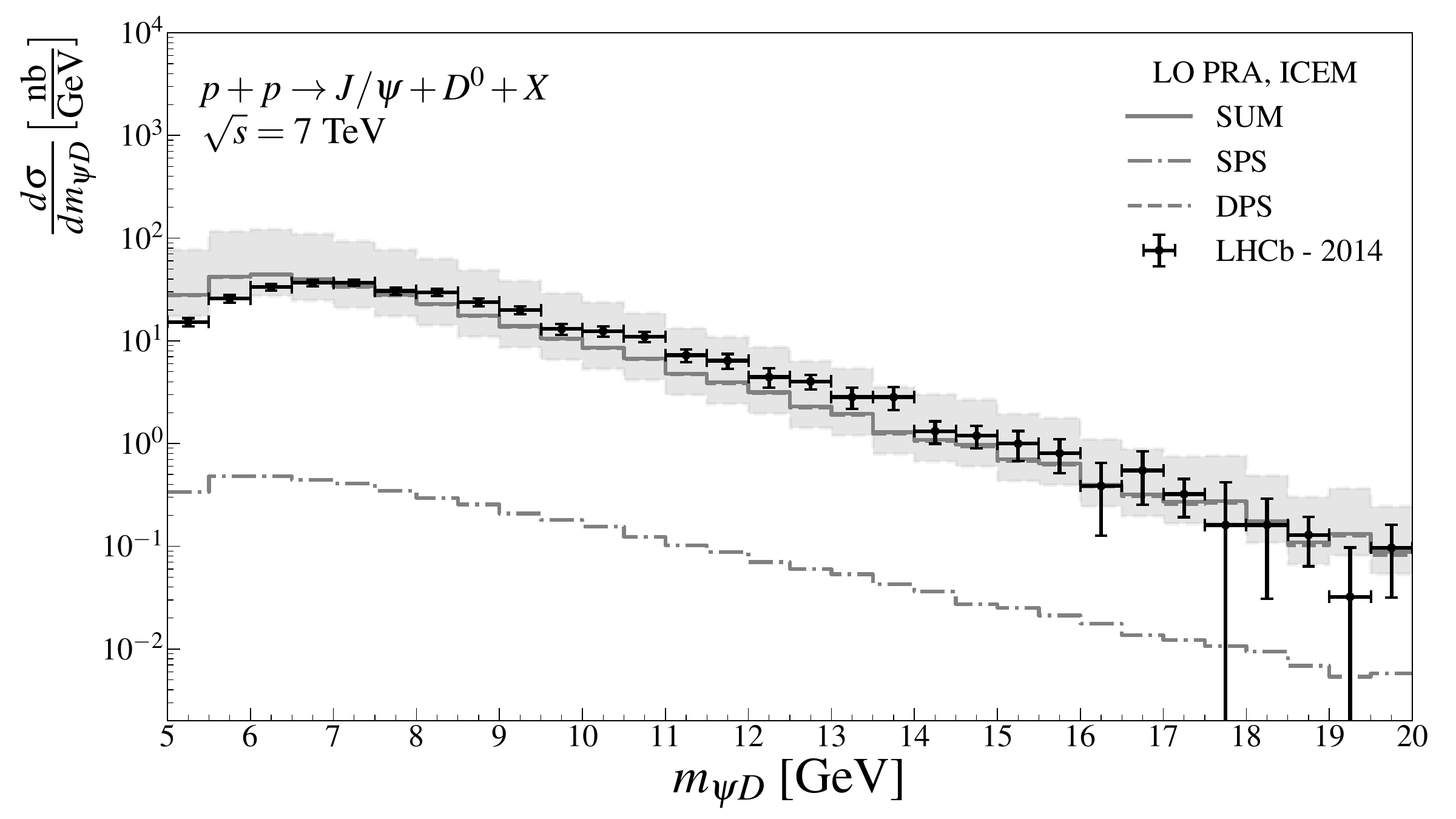}
\includegraphics[width=0.4\textwidth,angle=0]{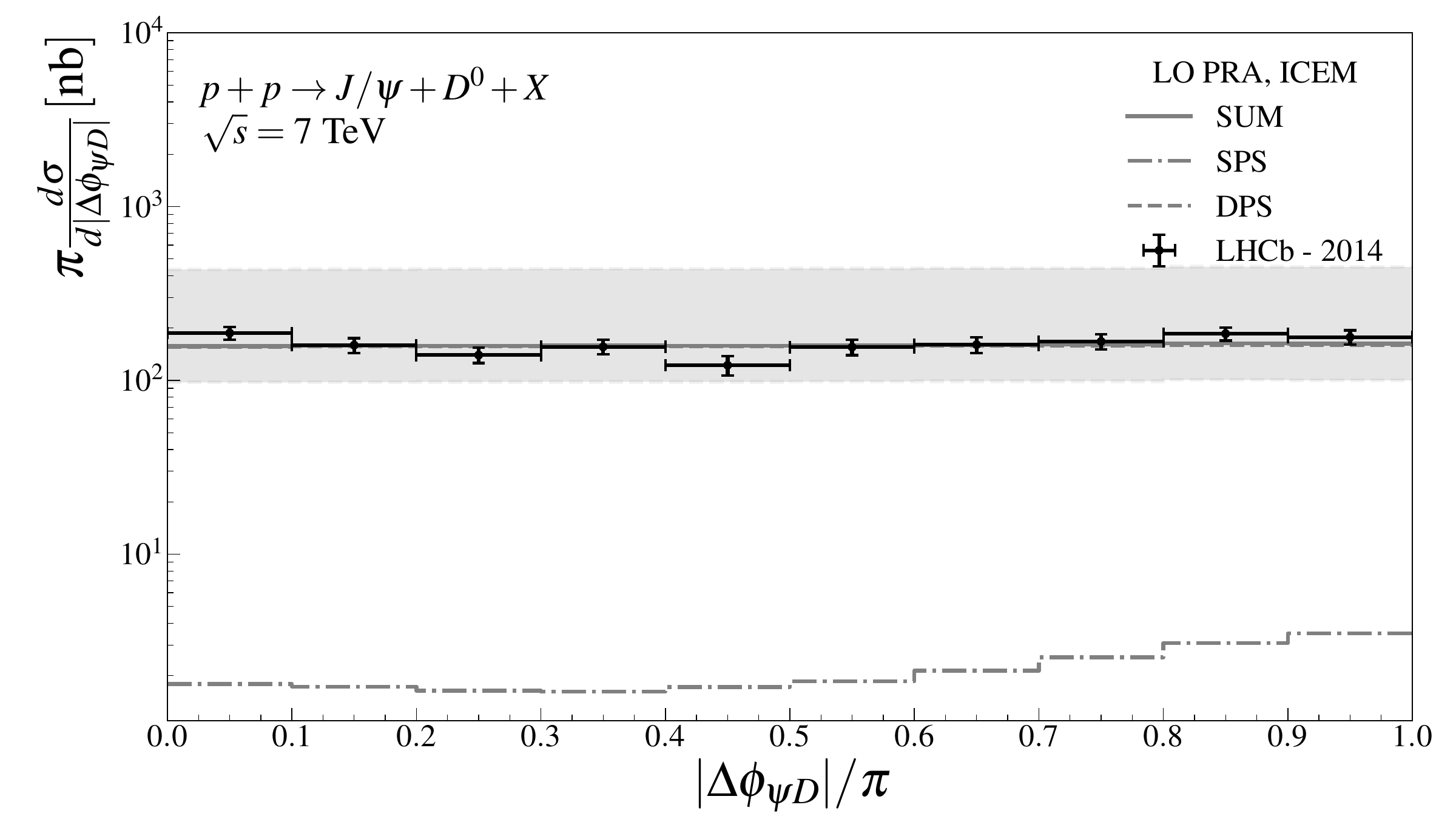}\includegraphics[width=0.4\textwidth,angle=0]{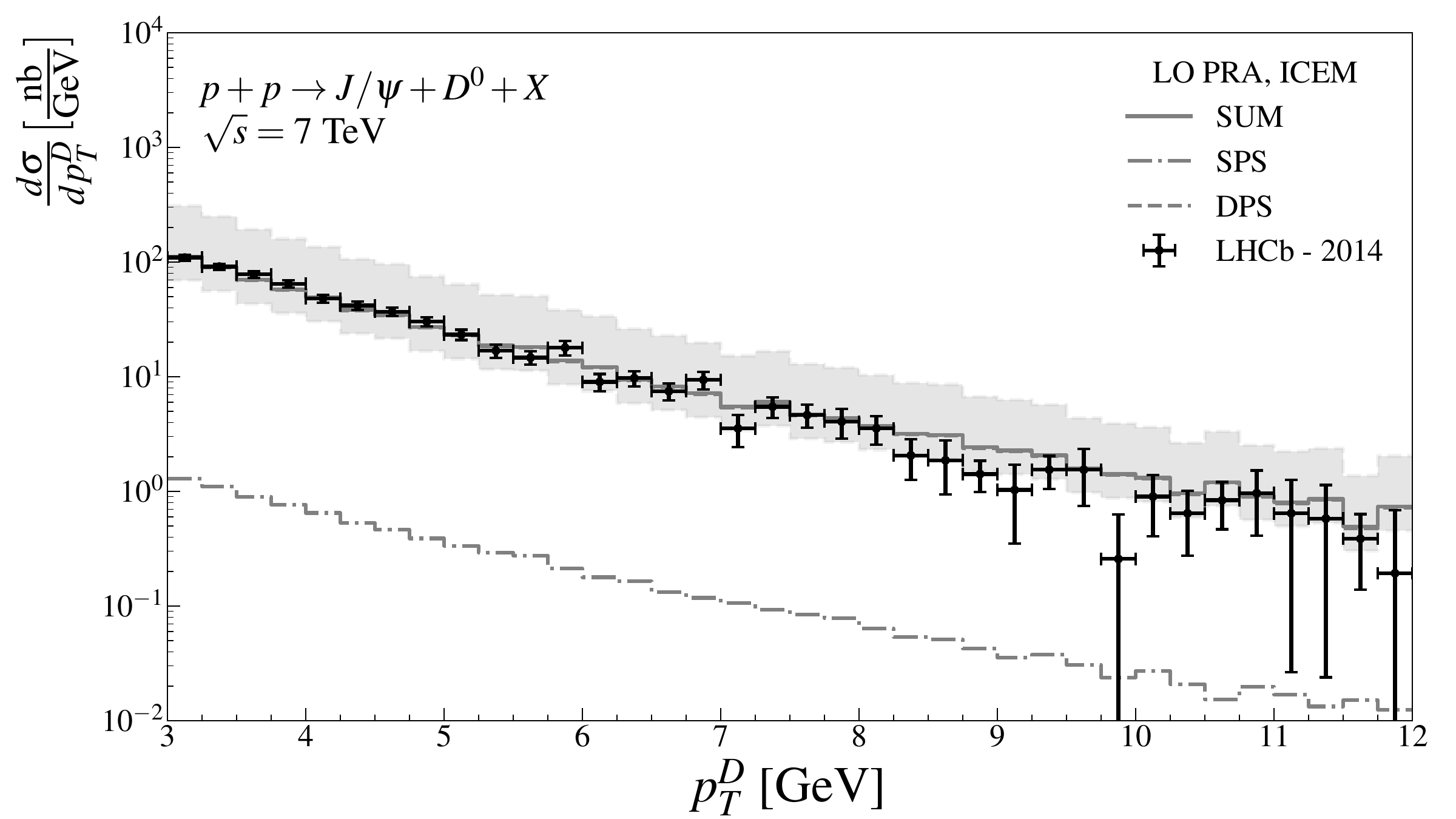}
\includegraphics[width=0.4\textwidth,angle=0]{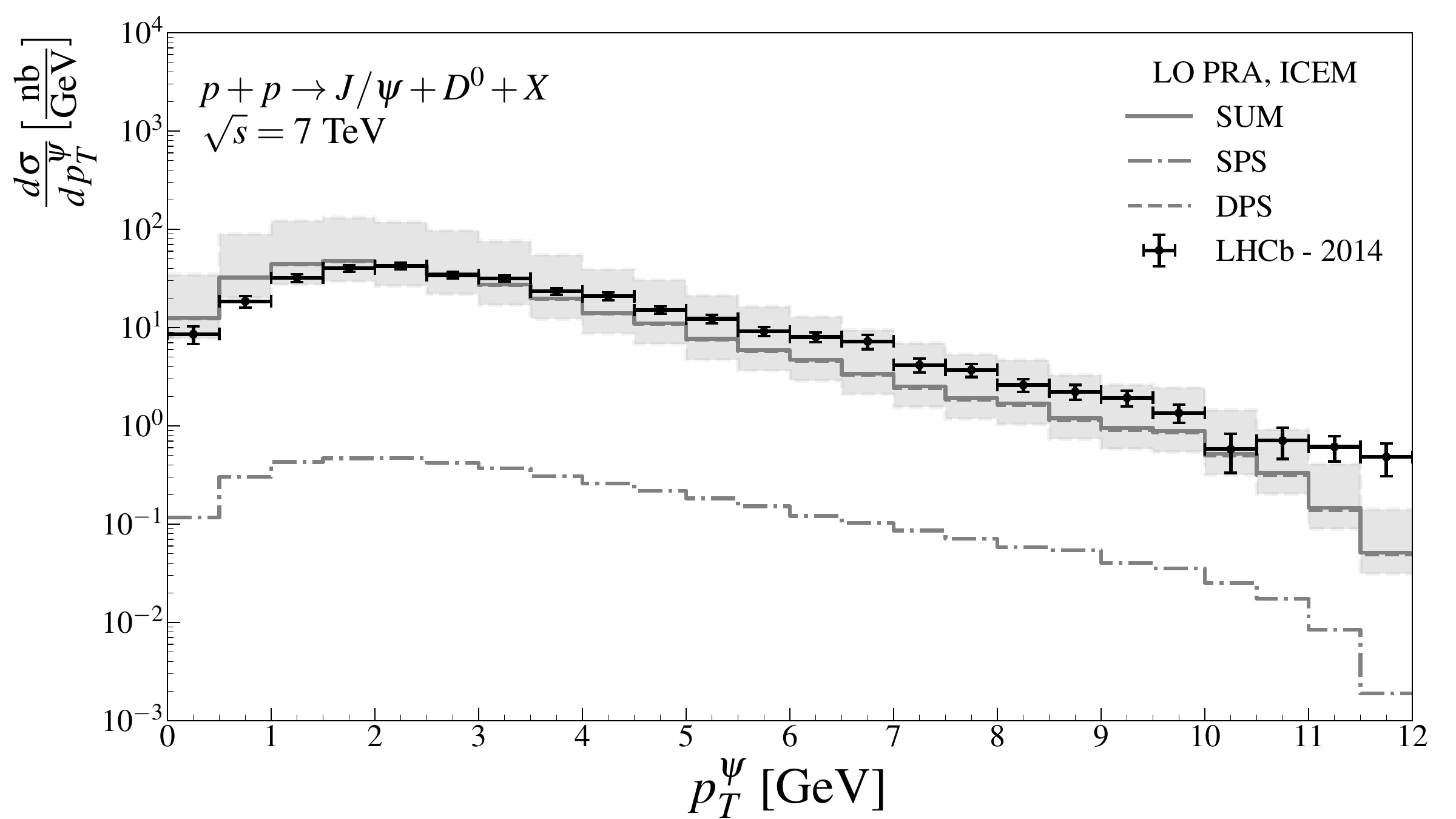}
\vspace{-3mm} \caption{ Various spectra for associated $J/\psi+D^0$
production at the $\sqrt{s}=7$ TeV: $\Delta y^{\psi D}$ - rapidity
difference, $m^{\psi D}$ - invariant mass, $|\Delta \phi^{\psi D}|$
- azimuthal angle difference, $p_T^D$ - $D^0$ transverse momentum
and $p_T^\psi$ - $J/\psi$ transverse momentum. The dashed line is
the DPS contribution, the dashed-dotted line is the SPS
contribution, solid line is they sum. Grey bounds around the solid
line are scale uncertainties of calculations. The data are from LHCb
collaboration \cite{LHCb:2012aiv} . \label{fig_6} }
\end{center}
\end{figure}

\begin{figure}[h]
\begin{center}
\includegraphics[width=0.4\textwidth,angle=0]{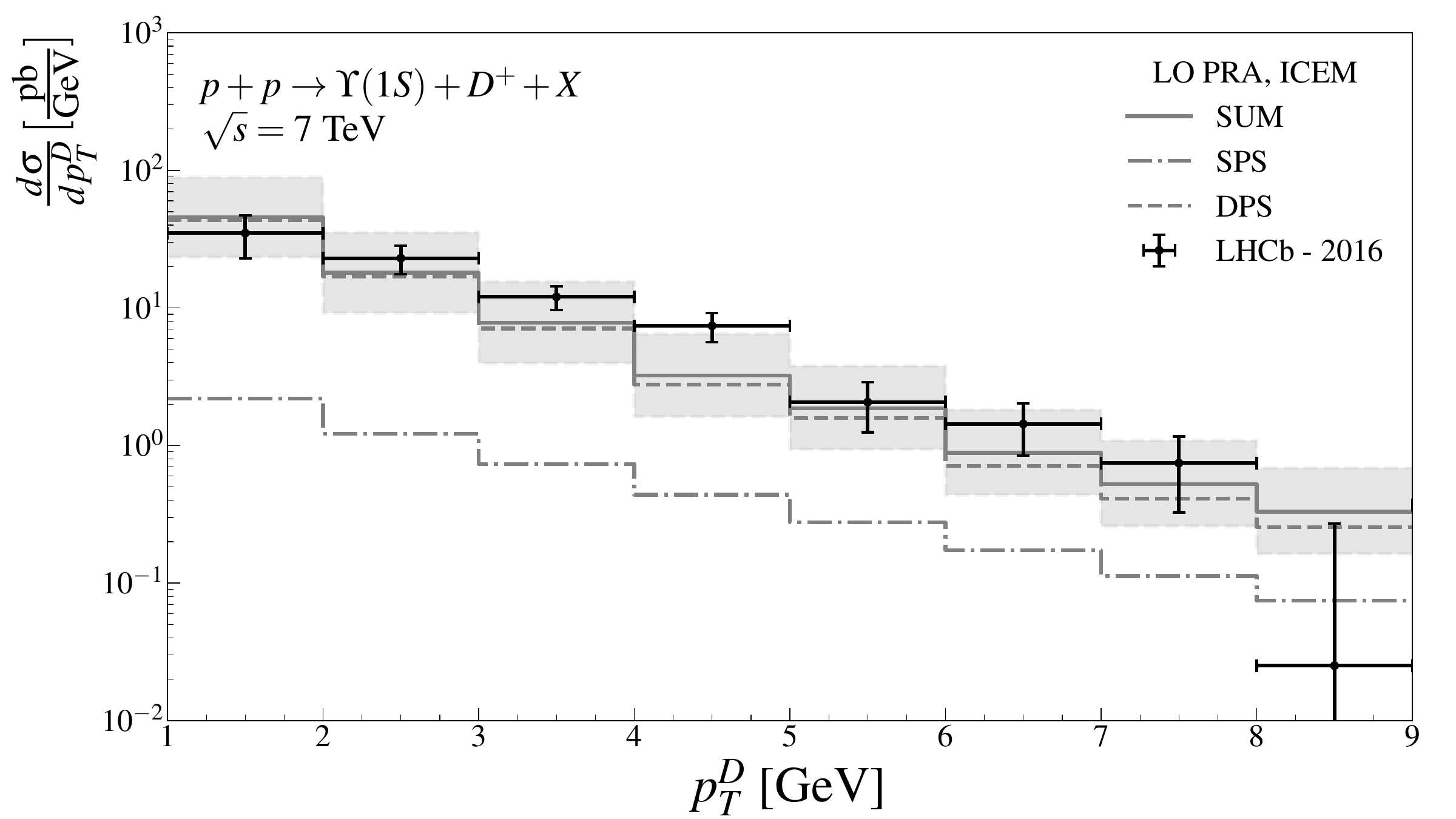}\includegraphics[width=0.4\textwidth,angle=0]{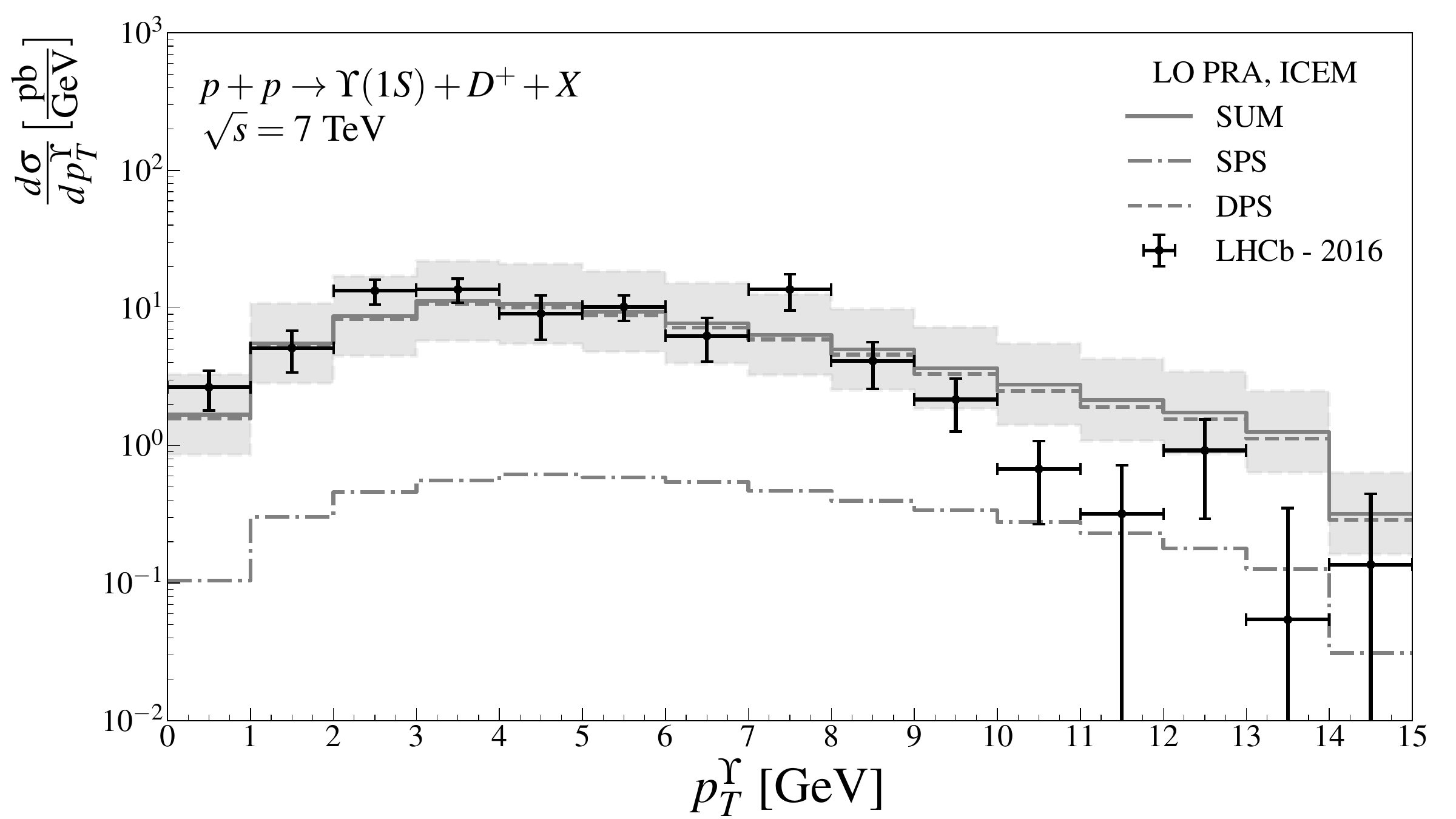}
\includegraphics[width=0.4\textwidth,angle=0]{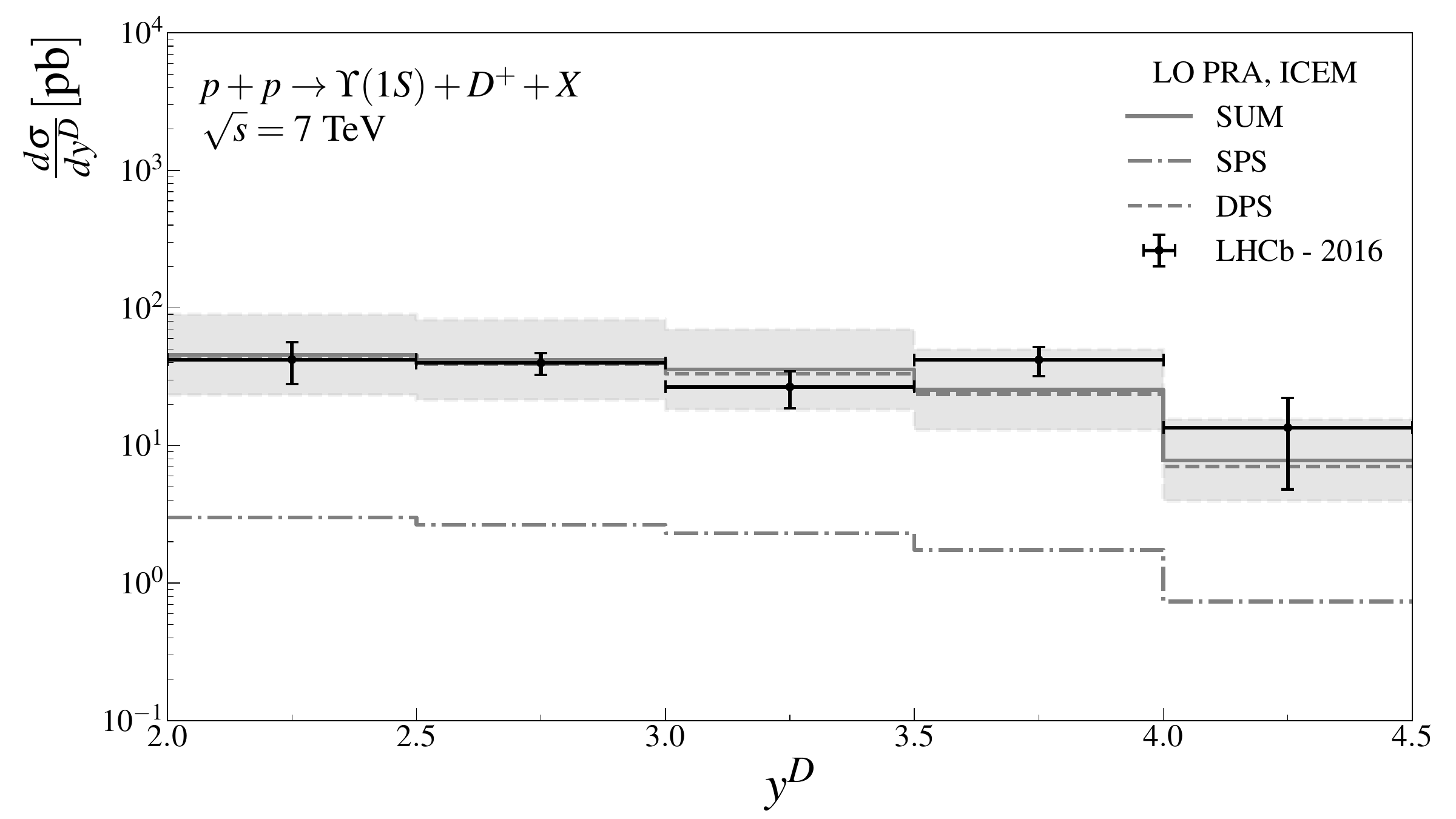}\includegraphics[width=0.4\textwidth,angle=0]{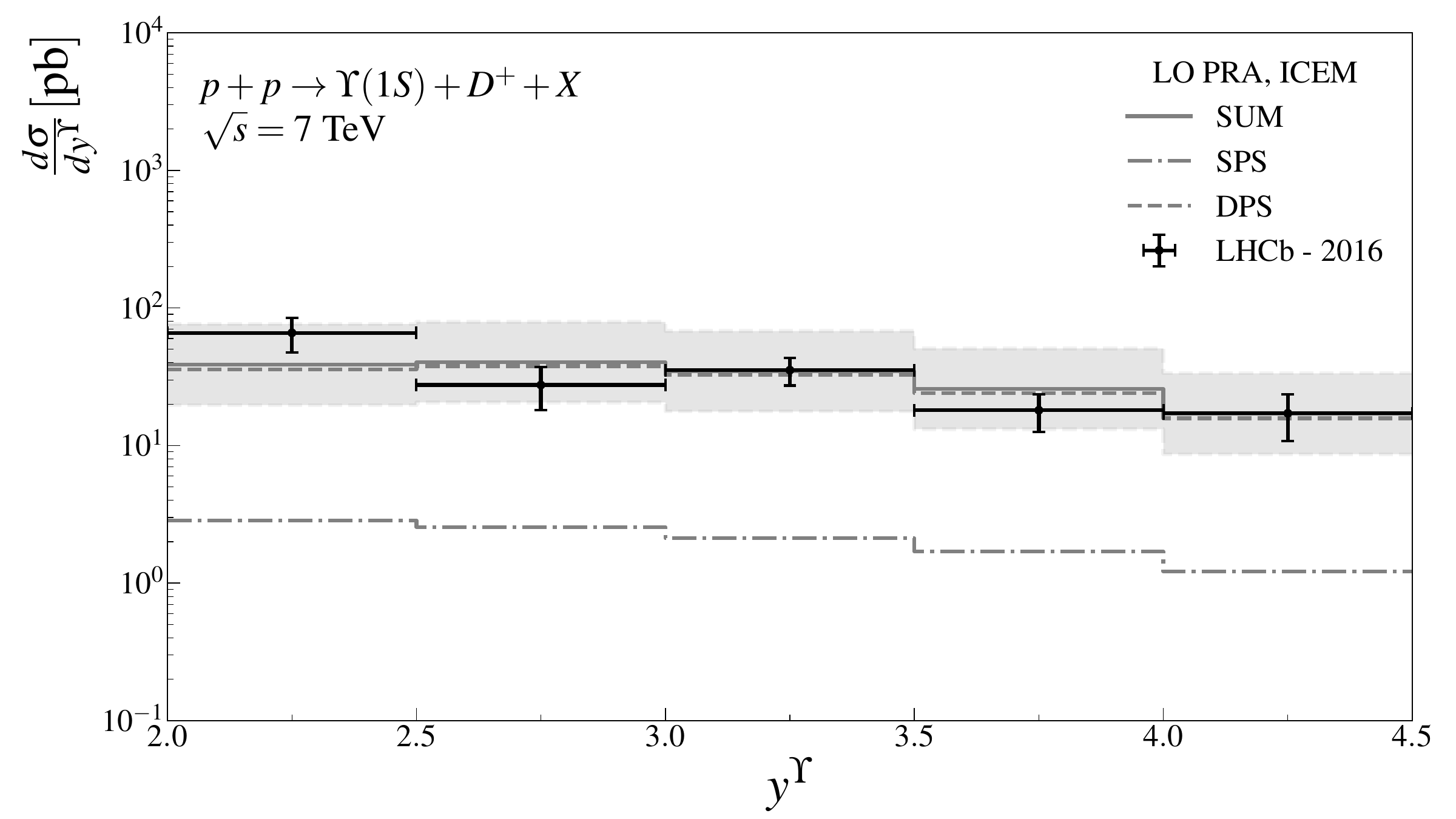}
\includegraphics[width=0.4\textwidth,angle=0]{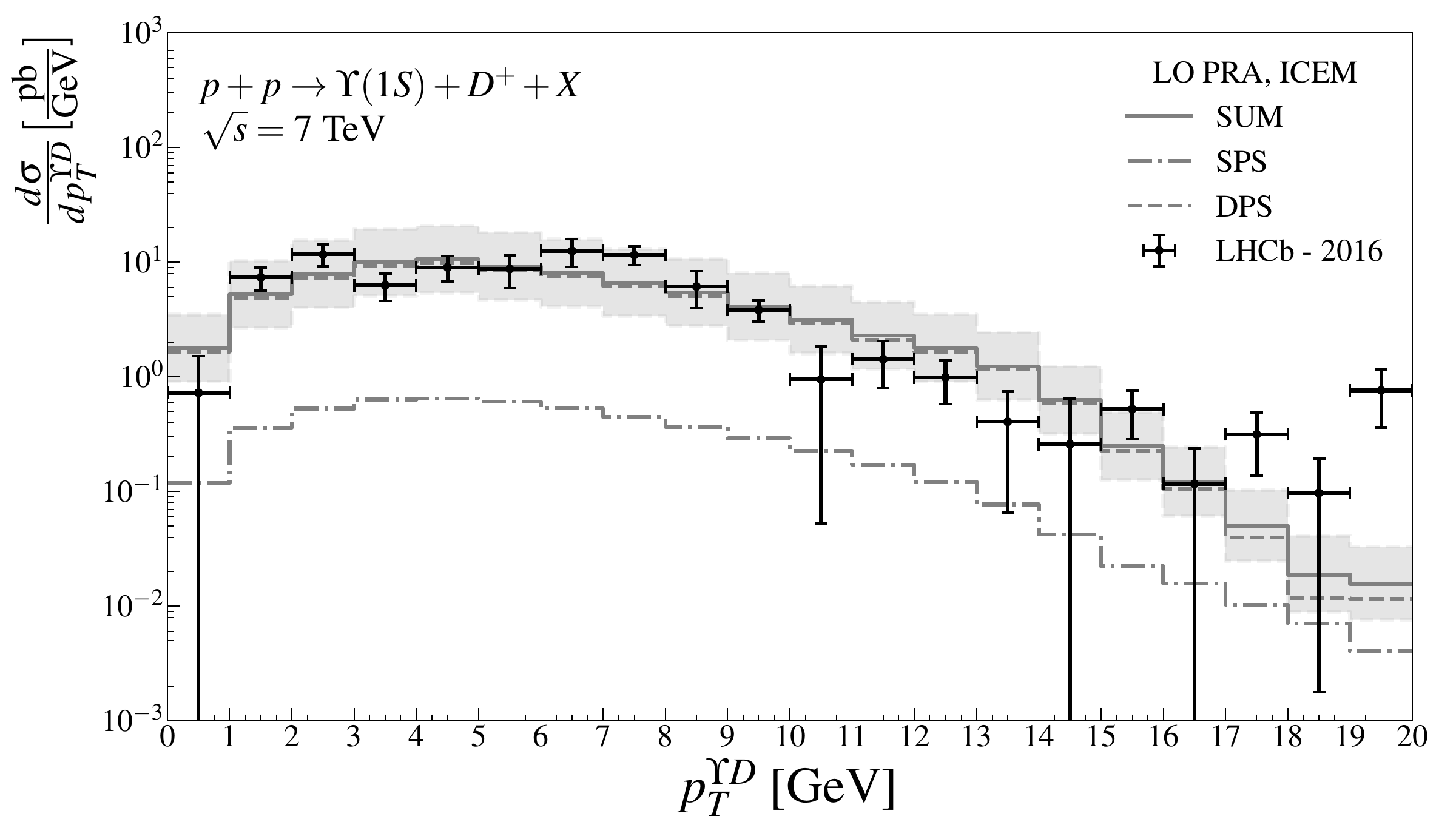}\includegraphics[width=0.4\textwidth,angle=0]{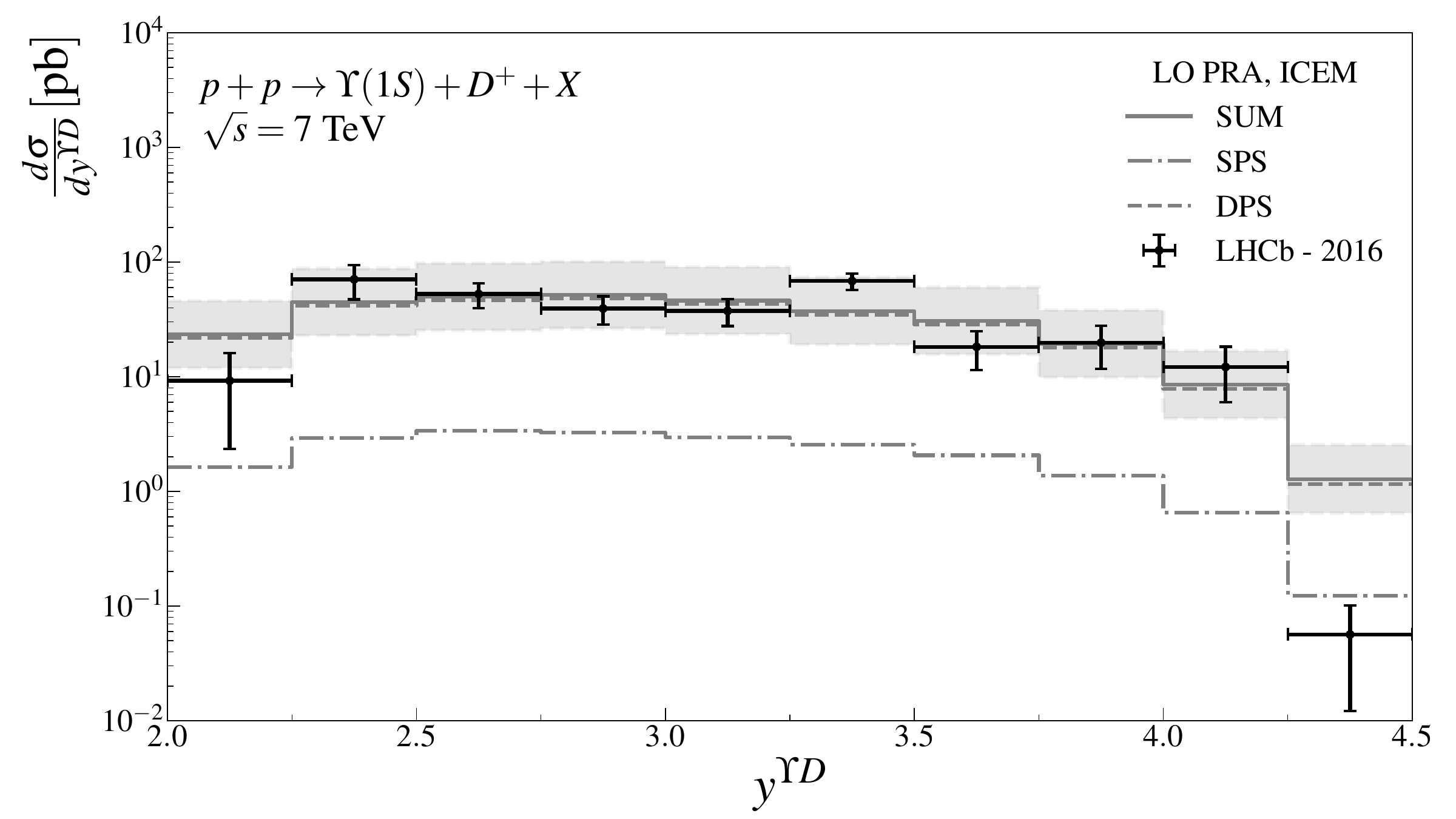}
\includegraphics[width=0.4\textwidth,angle=0]{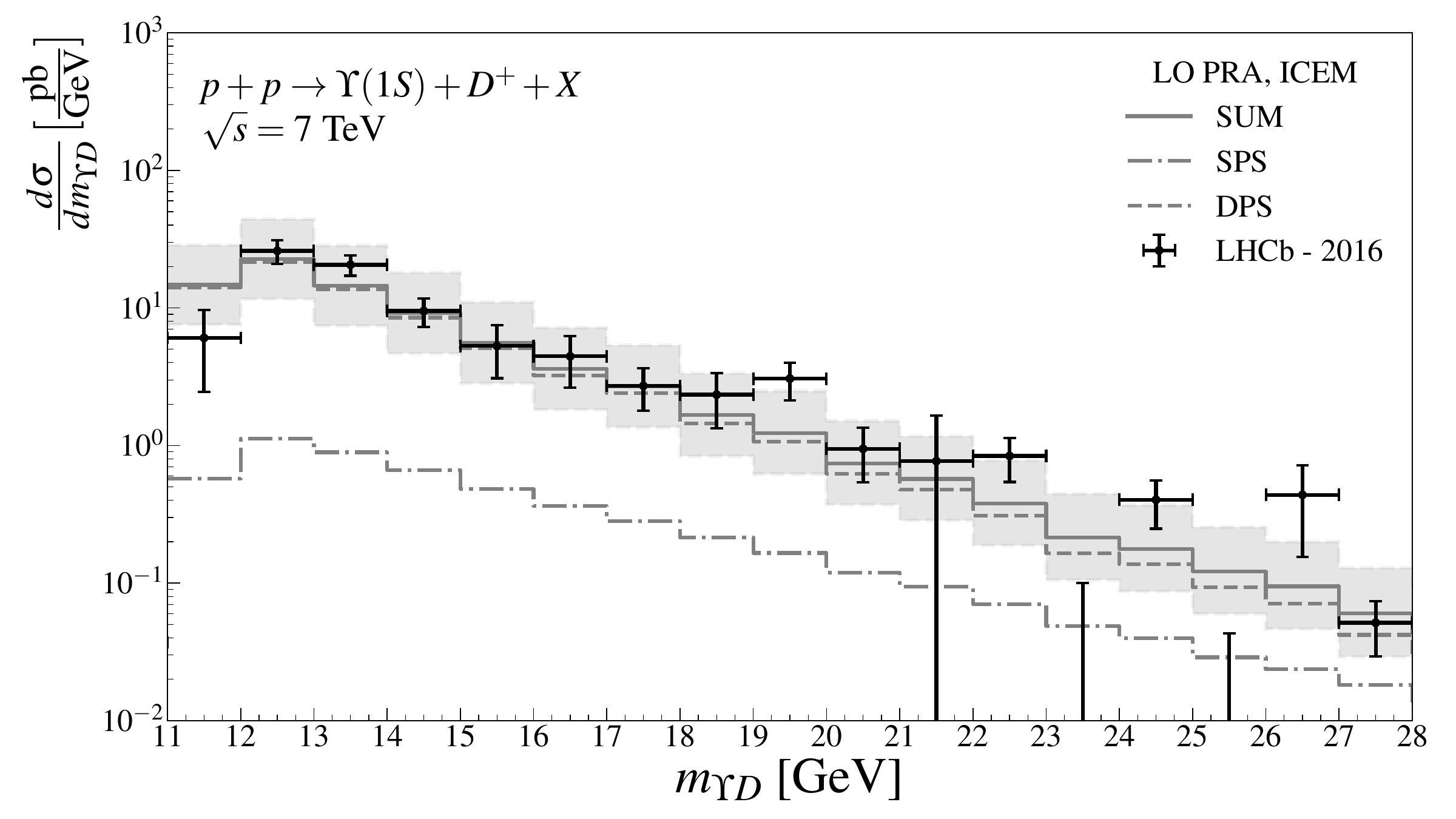}\includegraphics[width=0.4\textwidth,angle=0]{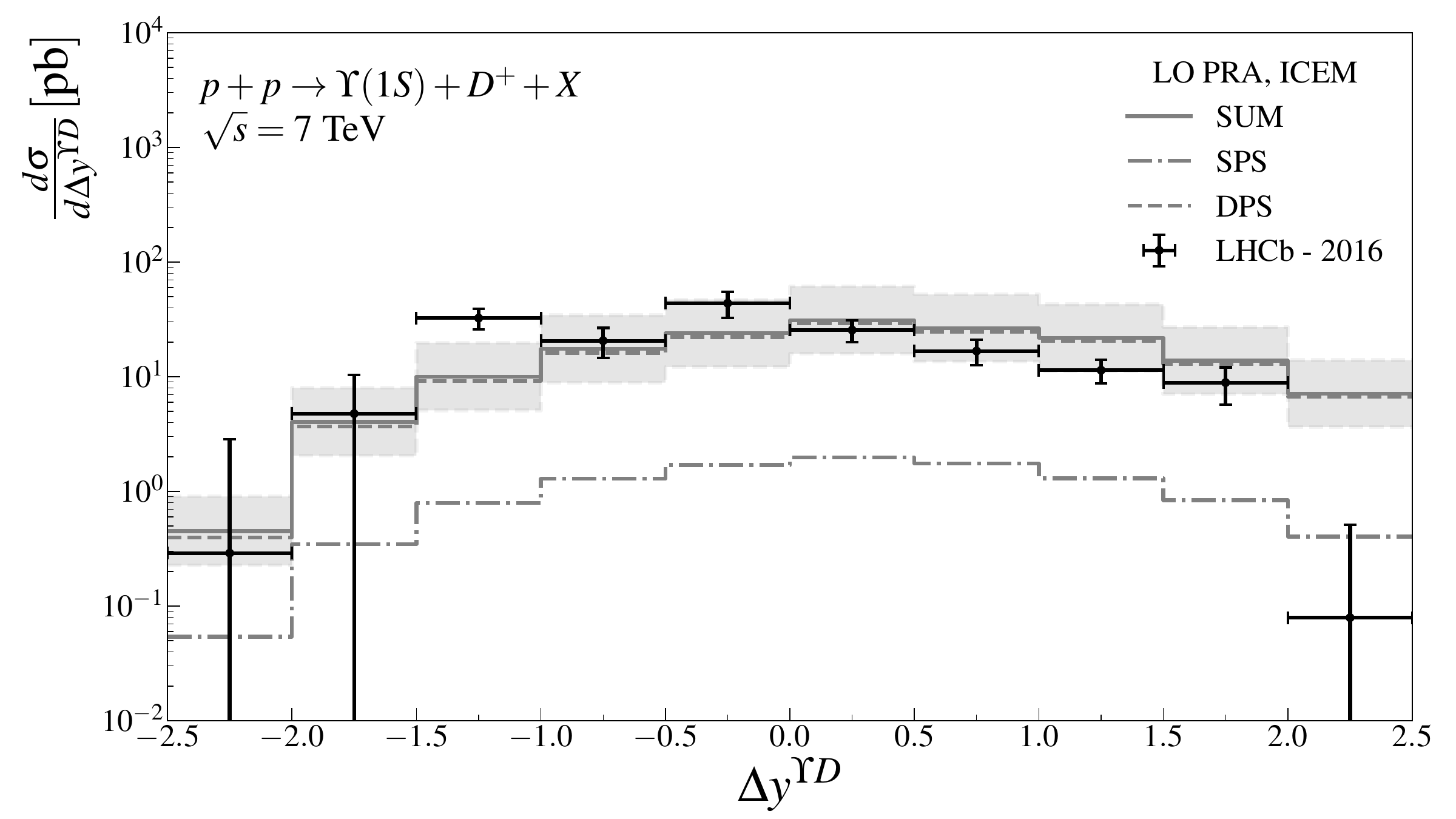}
\includegraphics[width=0.4\textwidth,angle=0]{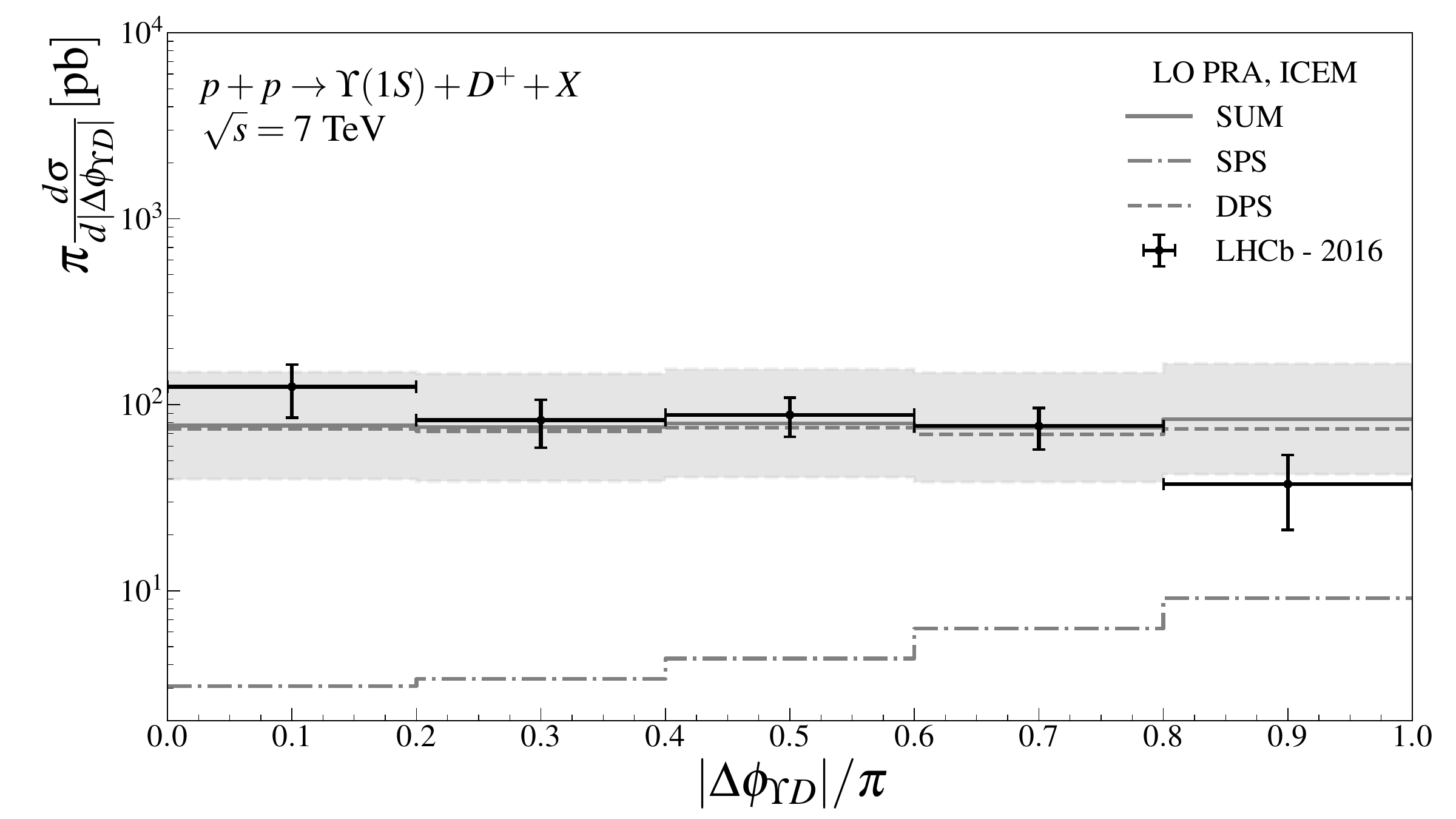}\includegraphics[width=0.4\textwidth,angle=0]{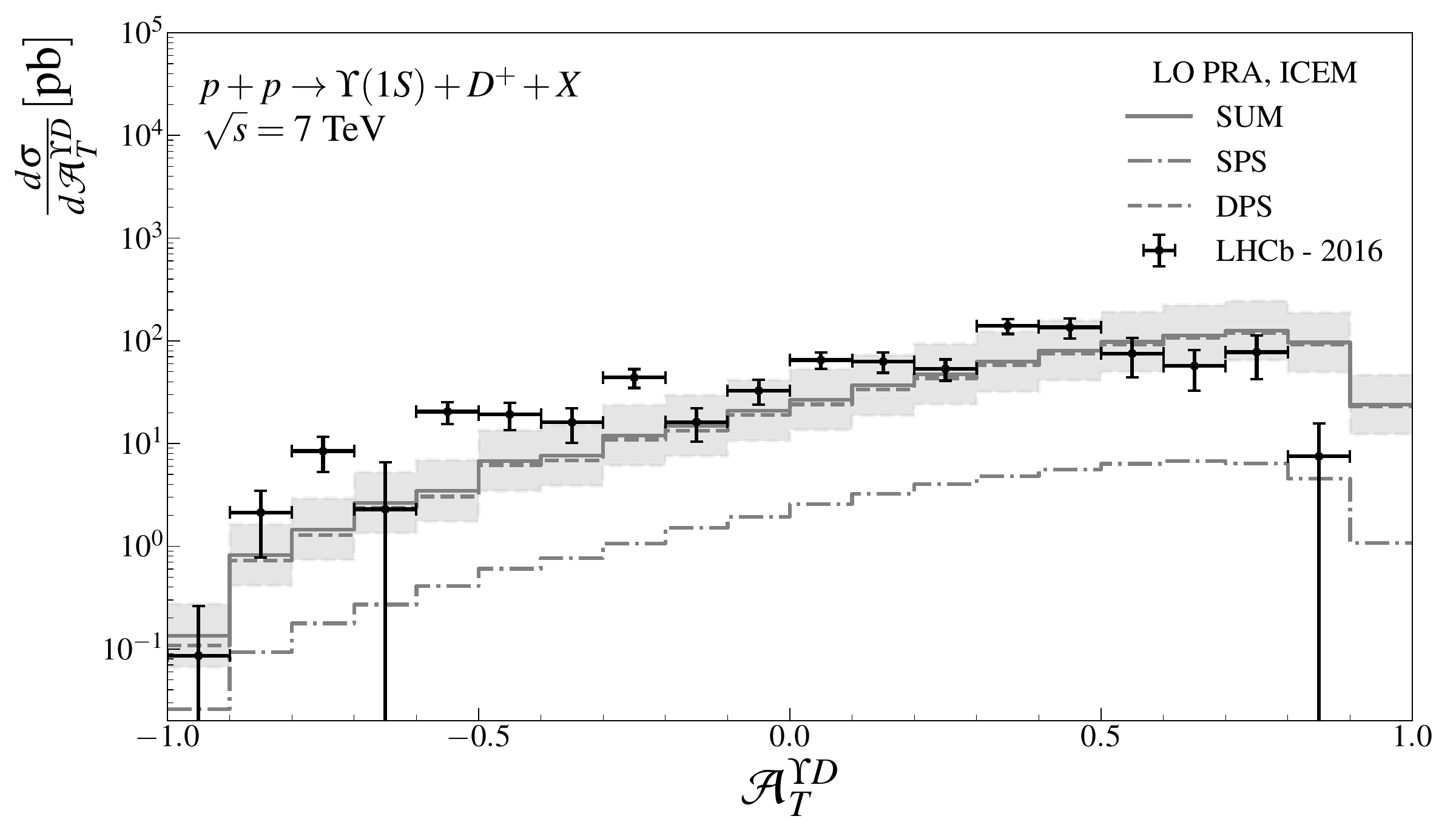}
\vspace{-3mm} \caption{ Various spectra for associated
$\Upsilon+D^+$ production at the $\sqrt{s}=7$ TeV: $y^\Upsilon$ -
$\Upsilon$ rapidity, $y^D$ - $D^+$ rapidity, $|\Delta y_{\Upsilon
D}|$ - rapidity difference, $m_{\Upsilon D}$ - invariant mass,
$|\Delta\phi_{\Upsilon D}|$ - azimuthal angle difference, $p_T^D$ -
$D^+$ transverse momentum, $p_T^\Upsilon$ - $\Upsilon$ transverse
momentum, $p_T^{\Upsilon D}$ - $\Upsilon +D^+$ transverse momentum,
$A_T^{\Upsilon D}$ - transverse momentum difference. The dashed line
is the DPS contribution, the dashed-dotted line is the SPS
contribution. Grey bounds around the solid line are scale
uncertainties of calculations.  The data are from LHCb collaboration
\cite{LHCb:2015wvu}. \label{fig_7} }
\end{center}
\end{figure}

\begin{figure}[h]
\begin{center}
\includegraphics[width=0.4\textwidth,angle=0]{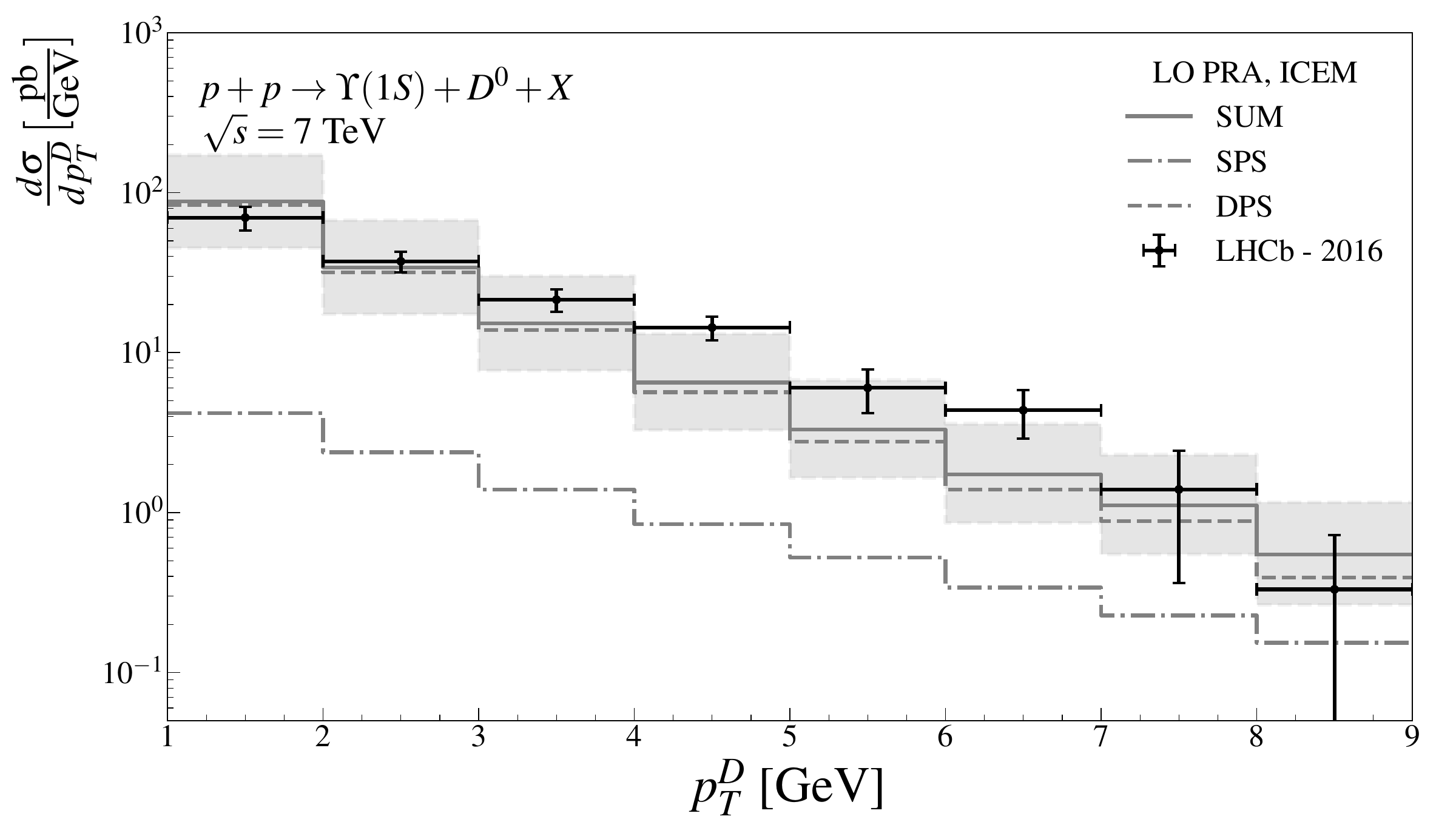}\includegraphics[width=0.4\textwidth,angle=0]{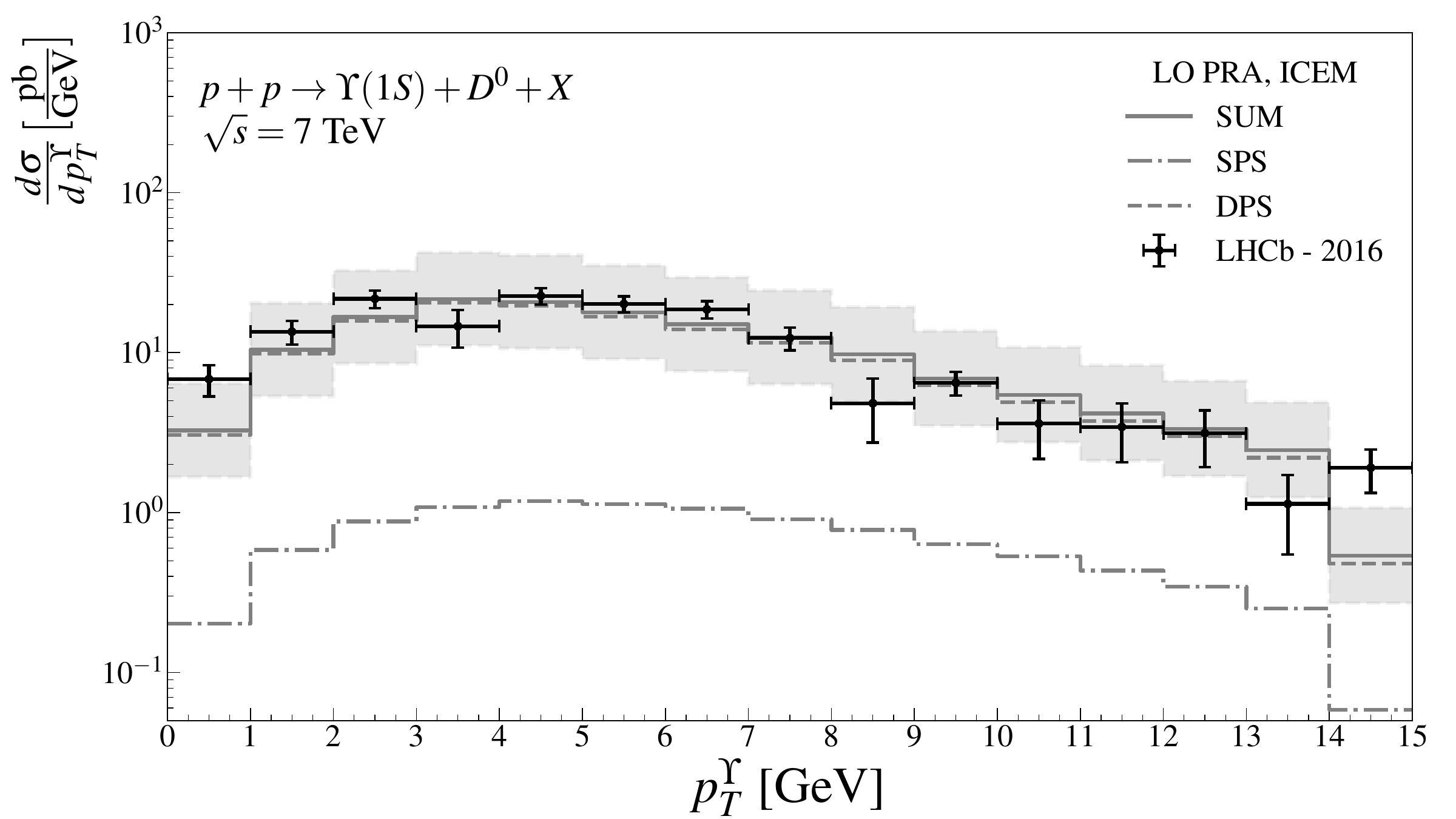}
\includegraphics[width=0.4\textwidth,angle=0]{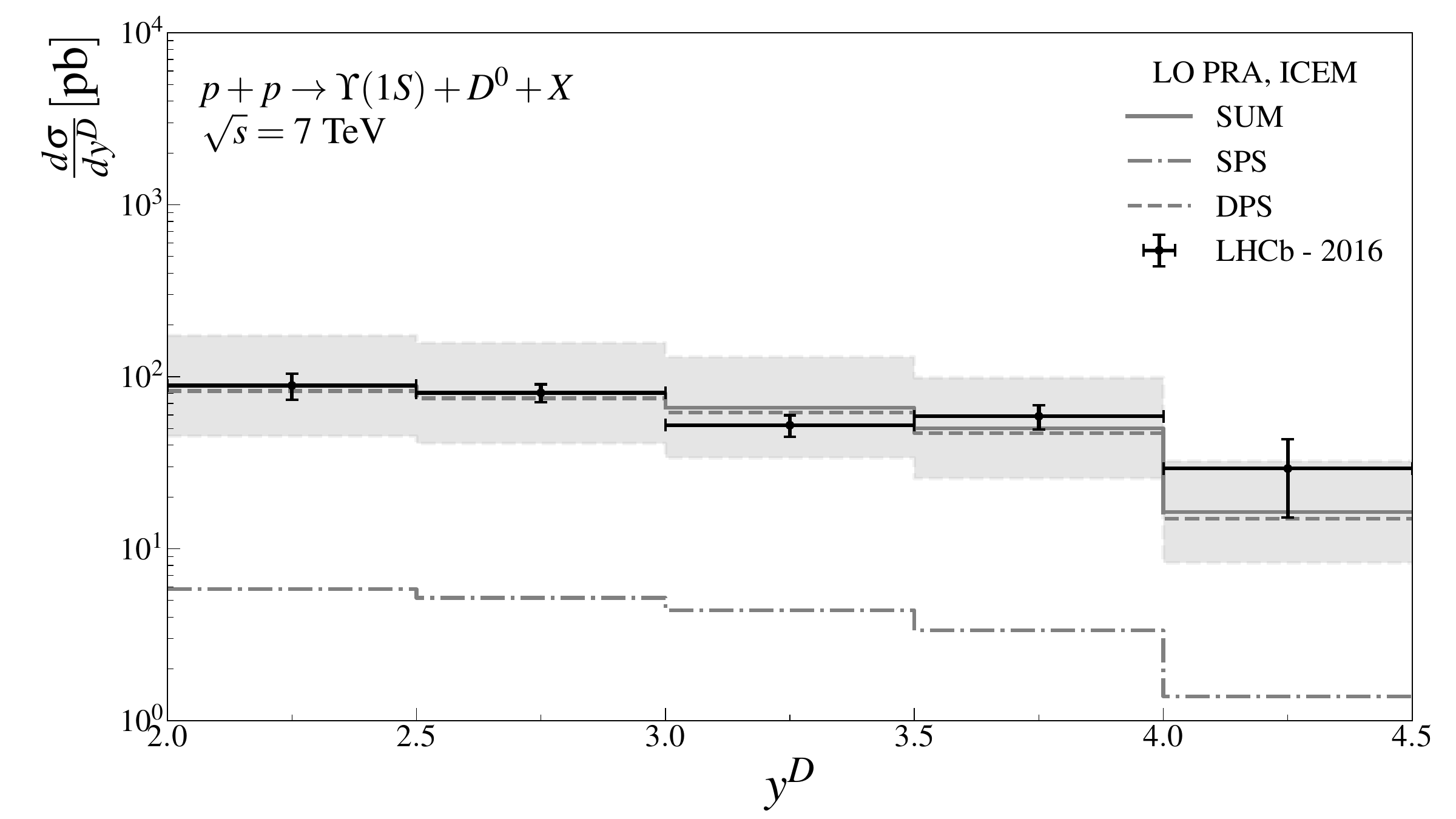}\includegraphics[width=0.4\textwidth,angle=0]{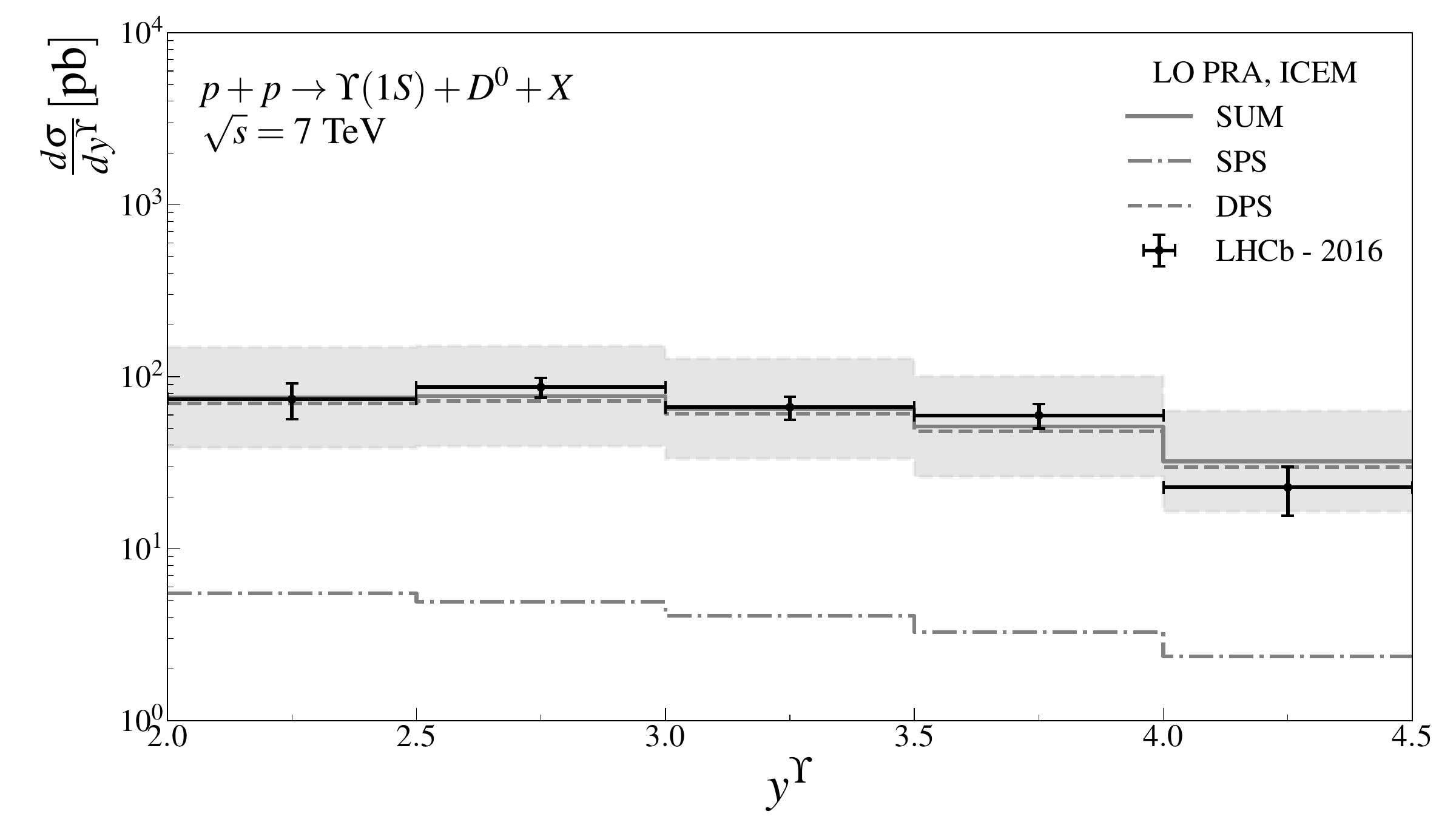}
\includegraphics[width=0.4\textwidth,angle=0]{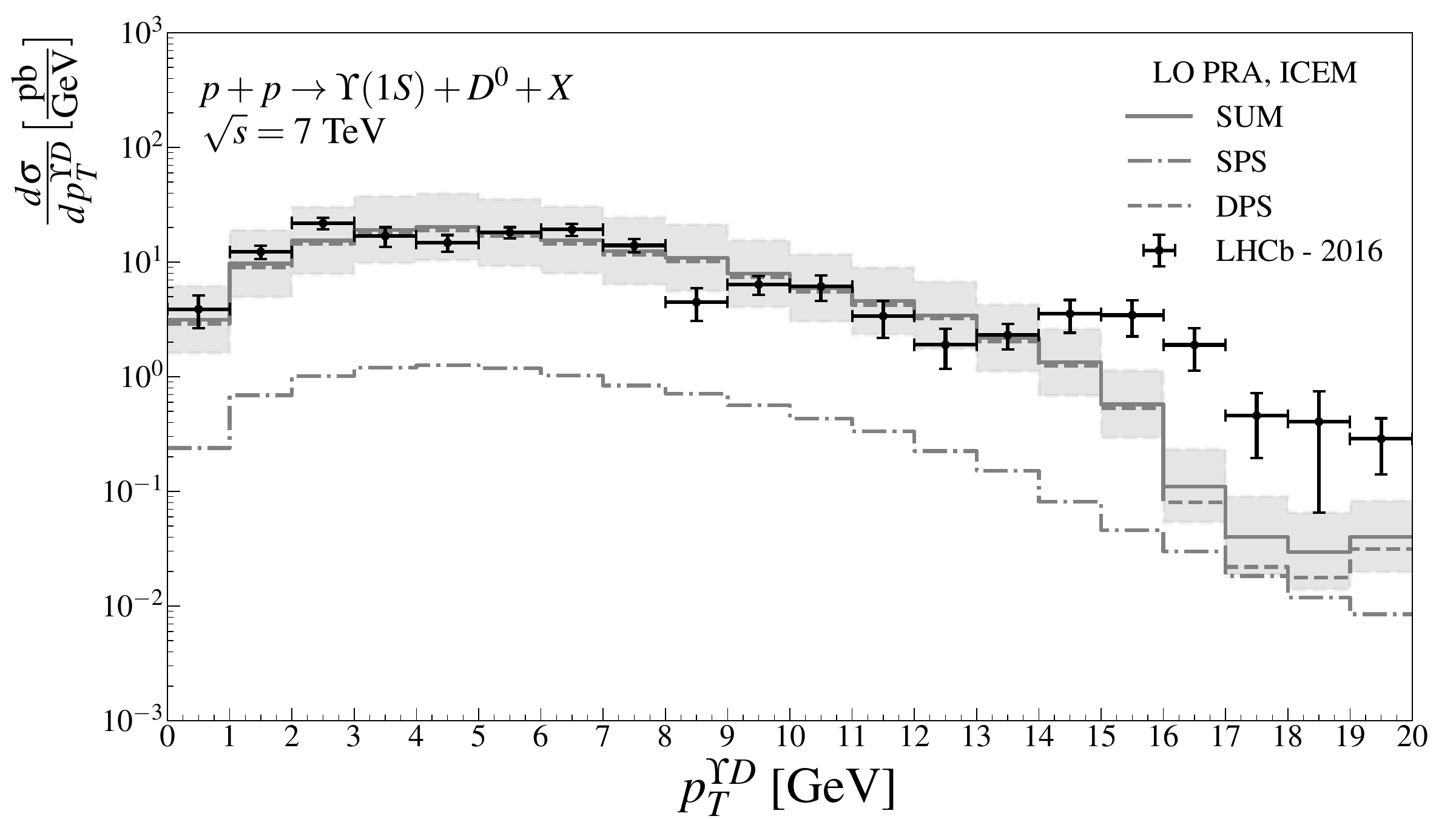}\includegraphics[width=0.4\textwidth,angle=0]{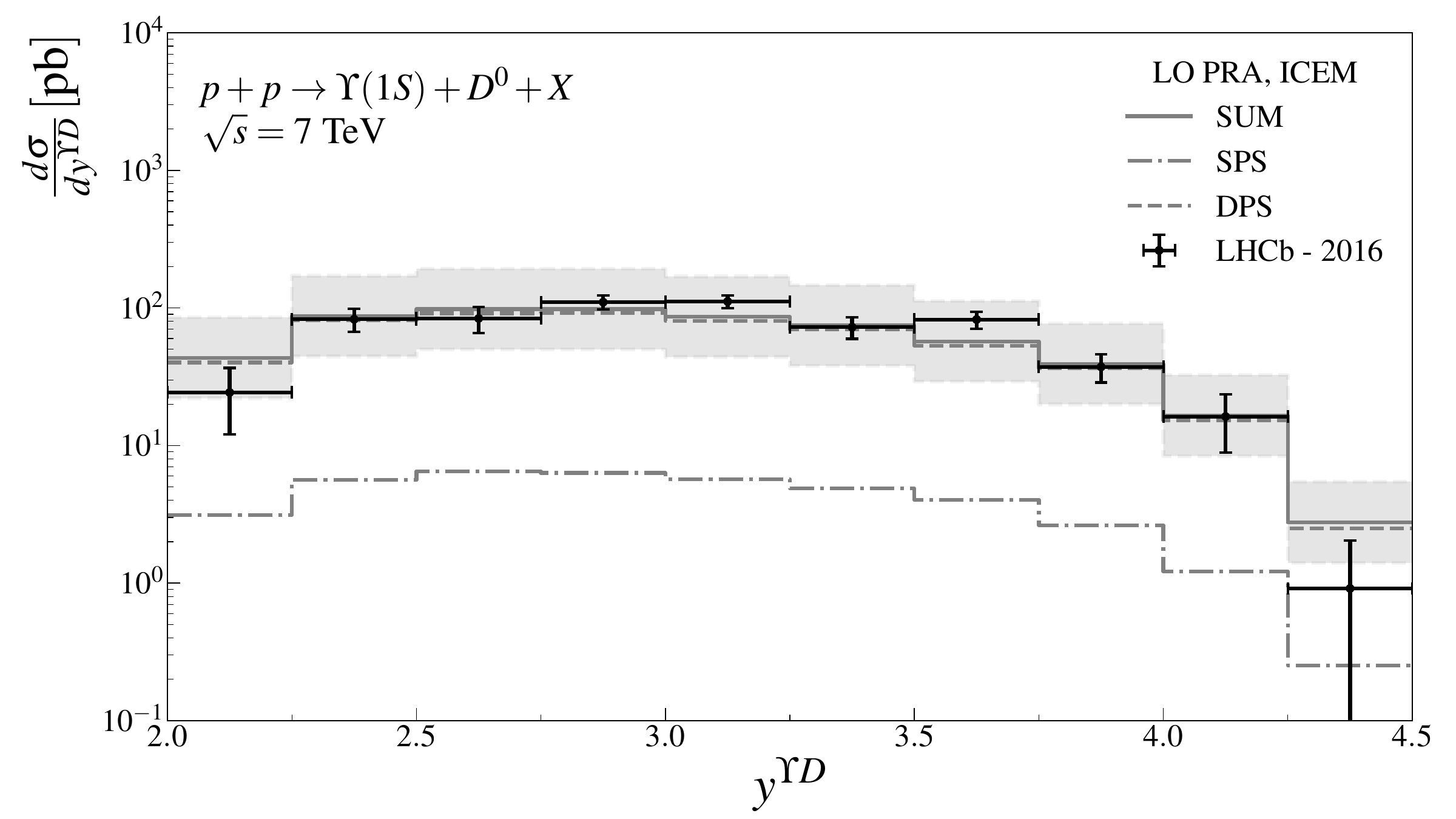}
\includegraphics[width=0.4\textwidth,angle=0]{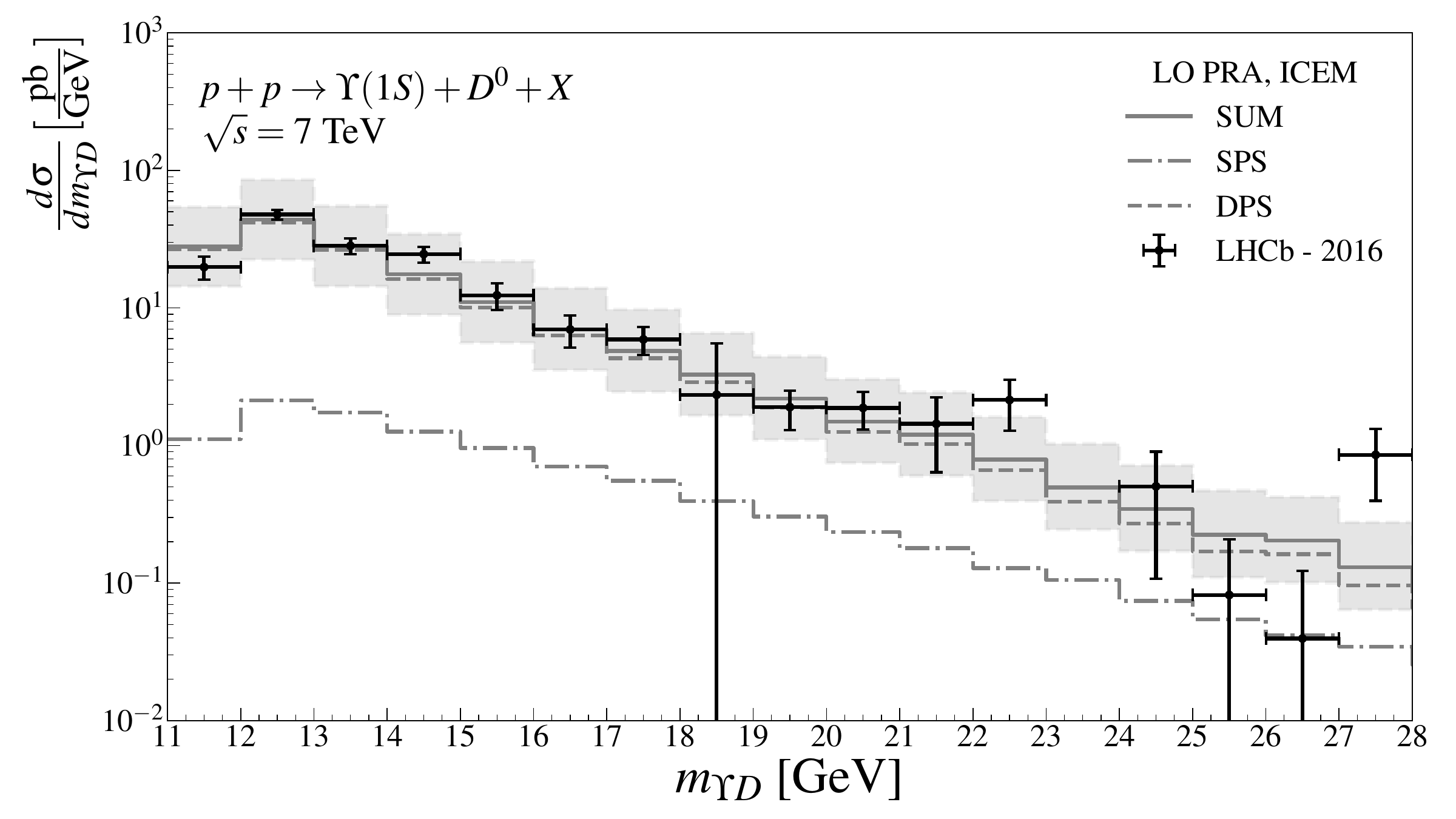}\includegraphics[width=0.4\textwidth,angle=0]{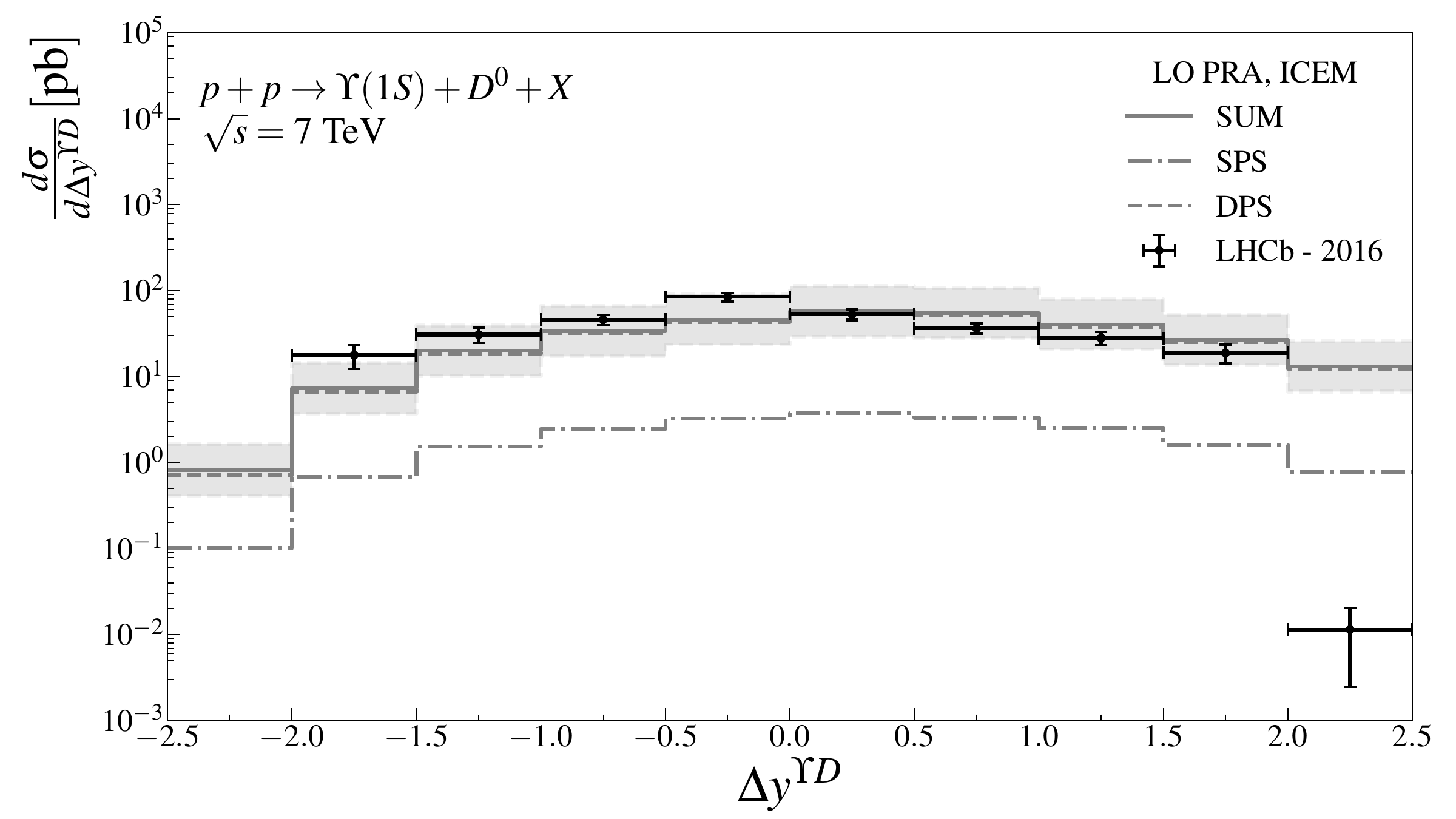}
\includegraphics[width=0.4\textwidth,angle=0]{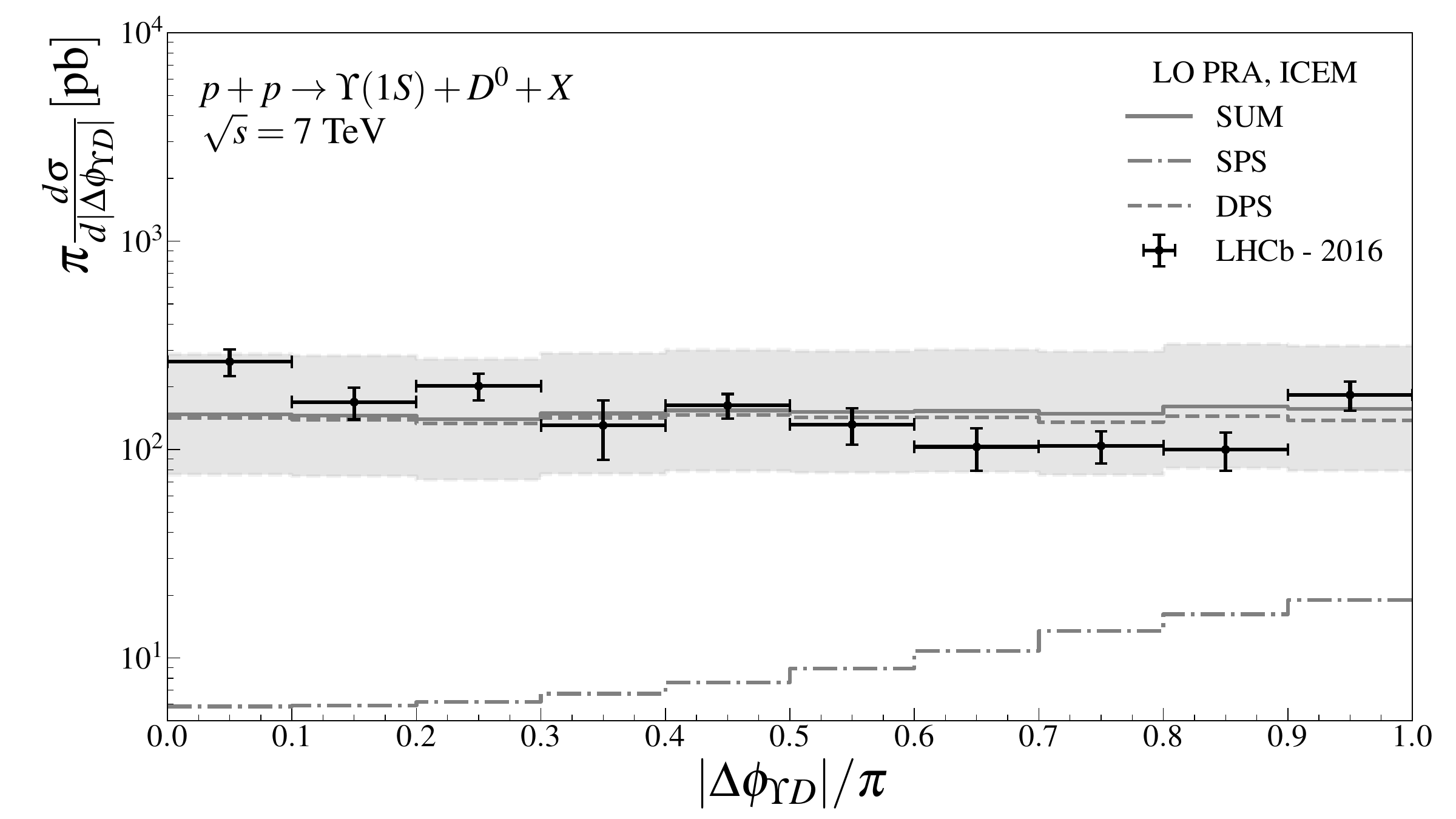}\includegraphics[width=0.4\textwidth,angle=0]{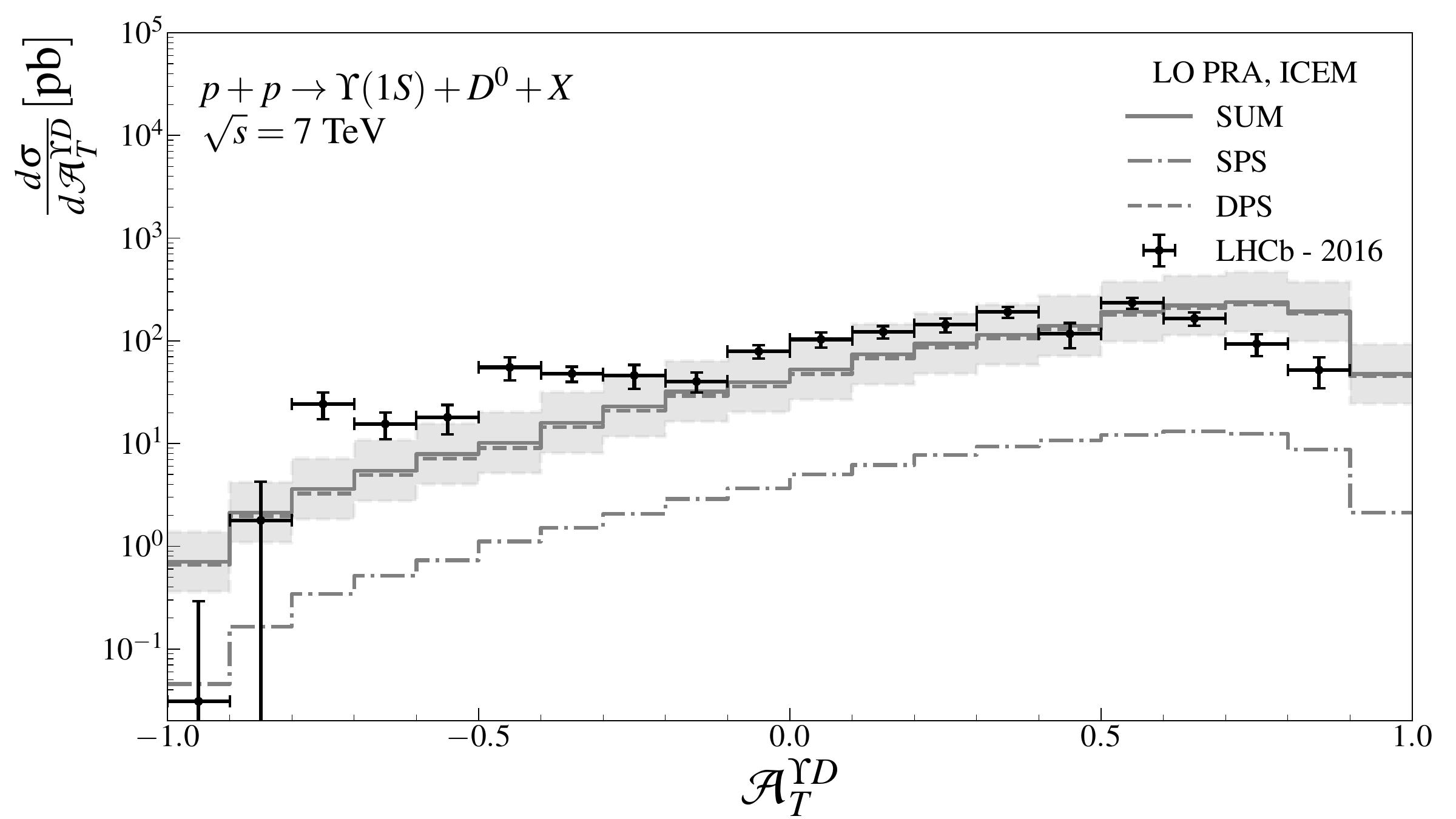}
\vspace{-3mm} \caption{ Various spectra for associated
$\Upsilon+D^0$ production at the $\sqrt{s}=7$ TeV: $y^\Upsilon$ -
$\Upsilon$ rapidity, $y^D$ - $D^0$ rapidity, $|\Delta y^{\Upsilon
D}|$ - rapidity difference, $m_{\Upsilon D}$ - invariant mass,
$|\Delta\phi_{\Upsilon D}|$ - azimuthal angle difference, $p_T^D$ -
$D^+$ transverse momentum, $p_T^\Upsilon$ - $\Upsilon$ transverse
momentum, $p_T^{\Upsilon D}$ - $\Upsilon +D^0$ transverse momentum,
$A_T^{\Upsilon D}$ - transverse momentum difference. The dashed line
is the DPS contribution, the dashed-dotted line is the SPS
contribution, solid line is they sum. Grey bounds around the solid
line are scale uncertainties of calculations. The data are from LHCb
collaboration \cite{LHCb:2015wvu}. \label{fig_8} }
\end{center}
\end{figure}

    \begin{table}[ht]
    \centering
\begin{tabular}{l c c r r}

\hline\hline Final state & Energy & Cross section & Experiment $\pm
\, (\rm stat) \pm (\rm syst)$ &
KaTie $\pm \, \Delta_{\rm SPS} \pm \Delta_{\rm DPS}$ \\
\hline

$J/\psi + D^0$ & $\sqrt{s} = 7$ TeV & ${\cal B}(J/\psi \to \mu^+
\mu^-) \times \sigma$ & $9.7 \pm 0.2 \pm 0.7 \ [\rm nb]$ &
$9.6 \ {}^{+ 0.4}_{- 0.1} \ {}^{+ 26.1}_{- 5.9} \ [\rm nb]$ \\

$J/\psi + D^+$ & $\sqrt{s} = 7$ TeV & ${\cal B}(J/\psi \to \mu^+
\mu^-) \times \sigma$ & $3.4 \pm 0.1 \pm 0.4 \ [\rm nb]$ &
$3.9 \ {}^{+ 0.2}_{- 0.02} \ {}^{+ 10.8}_{- 2.4} \ [\rm nb]$ \\

\hline

$\Upsilon + D^0$ & $\sqrt{s} = 7$ TeV & ${\cal B}(\Upsilon \to \mu^+
\mu^-) \times \sigma$ & $155 \pm 21 \pm 7 \ [\rm pb]$ &
$145 \ {}^{+ 16}_{- 6} \ {}^{+ 124}_{- 65} \ [\rm pb]$ \\

$\Upsilon + D^+$ & $\sqrt{s} = 7$ TeV & ${\cal B}(\Upsilon \to \mu^+
\mu^-) \times \sigma$ & $82 \pm 19 \pm 5 \ [\rm pb]$ &
$78 \ {}^{+ 14}_{- 2} \ {}^{+ 140}_{- 38} \ [\rm pb]$ \\

$\Upsilon + D^0$ & $\sqrt{s} = 8$ TeV & ${\cal B}(\Upsilon \to \mu^+
\mu^-) \times \sigma$ & $250 \pm 28 \pm 11 \ [\rm pb]$ &
$255 \ {}^{+ 25}_{- 9} \ {}^{+ 189}_{- 113} \ [\rm pb]$ \\

$\Upsilon + D^+$ & $\sqrt{s} = 8$ TeV & ${\cal B}(\Upsilon \to \mu^+
\mu^-) \times \sigma$ & $80 \pm 16 \pm 5 \ [\rm pb]$ &
$85 \ {}^{+ 8}_{- 3} \ {}^{+ 63}_{- 37} \ [\rm pb]$ \\

\hline

\multicolumn{5}{c}{Predictions} \\

\hline

$J/\psi + D^0$ & $\sqrt{s} = 13$ TeV & ${\cal B}(J/\psi \to \mu^+
\mu^-) \times \sigma$ & $-$ &
$34.3 \ {}^{+ 0.5}_{- 0.2} \ {}^{+ 36.9}_{- 17.6} \ [\rm nb]$ \\

$J/\psi + D^+$ & $\sqrt{s} = 13$ TeV & ${\cal B}(J/\psi \to \mu^+
\mu^-) \times \sigma$ & $-$ &
$14.2 \ {}^{+ 0.2}_{- 0.1} \ {}^{+ 15.3}_{- 7.3} \ [\rm nb]$ \\

$\Upsilon + D^0$ & $\sqrt{s} = 13$ TeV & ${\cal B}(\Upsilon \to
\mu^+ \mu^-) \times \sigma$ & $-$ &
$475 \ {}^{+ 31}_{- 12} \ {}^{+ 344}_{- 210} \ [\rm pb]$ \\

$\Upsilon + D^+$ & $\sqrt{s} = 13$ TeV & ${\cal B}(\Upsilon \to
\mu^+ \mu^-) \times \sigma$ & $-$ &
$197 \ {}^{+ 13}_{- 5} \ {}^{+ 143}_{- 82} \ [\rm pb]$ \\

\hline\hline
\end{tabular}
\caption{
    Comparison of theoretical and experimental total cross
    sections for associated  $J/\psi(\Upsilon) + D^{+,0}$ production.
    The theoretical uncertainties include the variation of the hard scale
    by factors $\xi = 2$, and $\frac{1}{2}$ from it's default  value
    $\mu = \frac{1}{2}\left( m^{\cal Q}_T + m^{D}_T \right)$.}
    \label{Table:1}

    \end{table}


\section{Conclusions}

Working in the ICEM and the $k_T-$factorization, taking into account
the SPS and the DPS mechanisms, we have obtained self agreement
description of the LHCb data for prompt heavy quarkonium
$(J/\psi,\Upsilon)$ production, inclusive $D^{+,0}-$meson
production, heavy quarkonium pair production $(J/\psi
J/\psi,\Upsilon\Upsilon, J/\psi \Upsilon)$
\cite{ChernyshevSaleev2Psi2022,ChernyshevSaleev2Ups2022}  and heavy
quarkonium plus $D^{+,0}-$meson associated production in the
high-energy proton-proton collisions. In this study, we confirm
early obtained numerical values for parameters of the ICEM, $F^\psi
\simeq F^\Upsilon \simeq 0.02$, and the DPS pocket formula,
$\sigma_{eff}\simeq 11$ mb, which don't contradict results obtained
previously by the different studies within the ICEM, the
$k_T-$factorization and the DPS model. It is shown that the
$k_T-$factorization, which involves into consideration high-order
QCD corrections included in uPDFs, may be a powerful tool to
calculate multi-particle production cross sections and spectra in
the multi-Regge kinematics. The efficiency of event generator KaTie
for calculations in the $k_T-$factorization approach is demonstrated
once more.

\section*{Acknowledgments}
We are grateful to A. van Hameren  for helpful communication on
 MC generator KaTie, A. Karpishkov, M. Nefedov and A. Shipilova for useful discussions.

\clearpage

\bibliography{references_D}
\end{document}